\documentclass[a4paper,11pt]{article}
\pdfoutput=1 
\usepackage{jcappub}

\usepackage{amsmath}
\usepackage{bm}
\usepackage{graphicx}
\usepackage{enumitem}
\usepackage{geometry}
\usepackage{xspace}
\usepackage{pdflscape}
\usepackage{placeins}
\usepackage{epstopdf}
\usepackage{multicol}

\makeatletter
\gdef\@fpheader{}
\g@addto@macro\bfseries{\boldmath}
\makeatother

\newcommand{\ie}{{i.e.~}}

\let\oldsqrt\sqrt
\def\sqrt{\mathpalette\DHLhksqrt}
\def\DHLhksqrt#1#2{%
\setbox0=\hbox{$#1\oldsqrt{#2\,}$}\dimen0=\ht0
\advance\dimen0-0.2\ht0
\setbox2=\hbox{\vrule height\ht0 depth -\dimen0}%
{\box0\lower0.4pt\box2}}

\newcommand{\dd}{\mathrm{d}}
\newcommand{\ee}{e}

\newcommand{\sss}[1]{{\scriptscriptstyle{#1}}}

\newcommand{\uPl}{\mathrm{Pl}}
\newcommand{\uin}{\mathrm{in}}

\newcommand{\uend}{\mathrm{end}}

\newcommand{\uini}{\mathrm{ini}}

\newcommand{\uc}{\mathrm{c}}

\newcommand{\usssPl}{\sss{\uPl}}

\newcommand{\Mp}{M_\usssPl}

\newcommand{\efolds}{$e$-folds}
\newcommand{\efold}{$e$-fold}

\newcommand{\beq}{\begin{eqnarray}}
\newcommand{\eeq}{\end{eqnarray}}
\newcommand{\bea}{\begin{equation}\begin{aligned}}
\newcommand{\eea}{\end{aligned}\end{equation}}

\newlength{\wsingfig}
\newlength{\wdblefig}
\newlength{\wquadfig}
\newlength{\wtriplefig}
\newcommand{\Eq}[1]{Eq.~(\ref{#1})}
\newcommand{\Eqs}[1]{Eqs.~(\ref{#1})}
\newcommand{\Fig}[1]{Fig.~{\ref{#1}}}
\newcommand{\Figs}[1]{Figs.~{\ref{#1}}}
\newcommand{\Ref}[1]{Ref.~{\cite{#1}}}
\newcommand{\Refs}[1]{Refs.~{\cite{#1}}}
\newcommand{\Sec}[1]{Sec.~\ref{#1}}

\subheader{}

\title{Metric preheating and radiative decay in single-field inflation}

\author[a]{J\'er\^ ome Martin,}
\author[b]{Theodoros Papanikolaou,}
\author[a]{Lucas Pinol,}
\author[b,a]{Vincent Vennin}

\affiliation[a]{Institut d'Astrophysique de Paris, UMR 7095-CNRS,
  Universit\'e Pierre et Marie Curie, 98bis boulevard Arago, 75014
  Paris, France}
\affiliation[b]{Laboratoire Astroparticule et
  Cosmologie, CNRS, Universit\'e de Paris, 75013 Paris, France}

\emailAdd{jmartin@iap.fr}
\emailAdd{theodoros.papanikolaou@apc.univ-paris7.fr}
\emailAdd{pinol@iap.fr}
\emailAdd{vincent.vennin@apc.univ-paris7.fr}

\date{today}

\begin{document}

\sloppy

\abstract{ At the end of inflation, the inflaton oscillates at the
  bottom of its potential and these oscillations trigger a parametric
  instability for scalar fluctuations with wavelength $\lambda$
  comprised in the instability band $(3H m)^{-1/2} <\lambda < H^{-1}$,
  where $H$ is the Hubble parameter and $m$ the curvature of the
  potential at its minimum. This ``metric preheating'' instability,
  which proceeds in the narrow resonance regime, leads to various
  interesting phenomena such as early structure formation, production
  of gravitational waves and formation of primordial black holes. In
  this work we study its fate in the presence of interactions with
  additional degrees of freedom, in the form of perturbative decay of
  the inflaton into a perfect fluid. Indeed, in order to ensure a
  complete transition from inflation to the radiation-dominated era,
  metric preheating must be considered together with perturbative
  reheating. We find that the decay of the inflaton does not alter the
  instability structure until the fluid dominates the universe
  content. As an application, we discuss the impact of the inflaton
  decay on the production of primordial black holes from the
  instability. We stress the difference between scalar field and
  perfect fluid fluctuations and explain why usual results concerning
  the formation of primordial black holes from perfect fluid
  inhomogeneities cannot be used, clarifying some recent statements
  made in the literature.}



\maketitle

\section{Introduction}
\label{sec:intro}

Cosmic inflation~\cite{Starobinsky:1980te, Guth:1980zm, Linde:1981mu,
  Albrecht:1982wi, Linde:1983gd} is presently the most promising
paradigm to describe the physical conditions that prevailed in the
very early universe. It consists of two stages. First, there is a
phase of accelerated expansion. In the simplest models, it is driven
by a scalar field, the inflaton, slowly rolling down its potential,
and the background spacetime almost exponentially expands. Second,
there is the reheating epoch~\cite{Albrecht:1982mp, Dolgov:1982th,
  Abbott:1982hn, Turner:1983he, Shtanov:1994ce, Kofman:1994rk,
  Kofman:1997yn} (see \Refs{Bassett:2005xm,Amin:2014eta} for reviews)
during which the inflaton field oscillates around the minimum of its
potential and decays into other degrees of freedom it couples
to. Then, after thermalisation of these decay products, the
radiation-dominated era of the hot big-bang phase starts.

One of the main successes of the inflationary scenario is that it
provides a convincing mechanism for the origin of the structures in
our universe~\cite{Mukhanov:1981xt, Kodama:1985bj}. According to the
inflationary paradigm, they stem from quantum fluctuations born on
sub-Hubble scales and subsequently amplified by gravitational
instability and stretched to super-Hubble distances by cosmic
expansion. During this process, which occurs in the slow-roll phase,
cosmological perturbations acquire an almost scale-invariant power
spectrum, which is known to provide an excellent fit to the
astrophysical data at our disposal~\cite{Akrami:2018vks,
  Akrami:2018odb}.

In the simplest models where inflation is driven by a single scalar
field with canonical kinetic term, on large scales, the curvature
perturbation is conserved~\cite{Mukhanov:1981xt, Kodama:1985bj}, which
implies that the details of the reheating process do not affect the
inflationary predictions or, in other words, that ``metric
preheating'' is inefficient on those scales. Since these models are
well compatible with the data~\cite{Martin:2013tda, Martin:2013nzq,
  Martin:2015dha, Chowdhury:2019otk}, the stage of reheating is
usually not considered as playing an important role in the evolution
of cosmological perturbations~\cite{Finelli:1998bu}. For the scales
observed in the Cosmic Microwave Background (CMB), the only effect of
reheating is through the amount of expansion that proceeds during this
epoch, which relates physical scales as we observe today to the time
during inflation when they emerge. This thus determines the part of
the inflationary potential that we probe with the CMB. In practice,
there is a single combination~\cite{Martin:2006rs} of the reheating
temperature and of the mean equation-of-state parameter, that sets the
location of the observational window along the inflationary
potential. Given the restrictions on the shape of the potential now
available~\cite{Martin:2013nzq, Vennin:2015eaa}, this can be used to
constrain the kinematics of reheating~\cite{Martin:2010kz,
  Martin:2014nya, Martin:2016oyk, Hardwick:2016whe}. In multiple-field
scenarios, on the contrary, large-scale curvature perturbations can be
strongly distorted by the reheating
process~\cite{Bassett:1999cg,Bassett:1999mt,
  Jedamzik:1999um,Finelli:2000ya,Allahverdi:2010xz}, which means that
metric preheating can be important and, thus, can have more impact on
CMB observations.

The situation is very different for scales smaller than those observed
in the CMB, more precisely for scales crossing back in the Hubble
radius during reheating (or never crossing out the Hubble radius
during inflation). In particular, it was shown in
\Ref{Jedamzik:2010dq} (see also \Ref{Easther:2010mr}) that the density
contrast of the scalar field fluctuations can grow on small scales
during preheating, due to a parametric instability sourced by the
oscillations of the inflaton at the bottom of its potential. This
mechanism demonstrates that metric preheating can be important even in
single-field inflation, although not on large scales. It can give rise
to different interesting phenomena such as early structure
formation~\cite{Jedamzik:2010dq}, gravitational waves
production~\cite{Jedamzik:2010hq} and even Primordial Black Holes
(PBHs)~\cite{Carr:1974nx,Carr:1975qj} formation~\cite{Martin:2019nuw}
(PBHs formation from scalar fields was considered in
\Ref{Khlopov:1985jw}, in the case of two-fields models in
\Ref{Bassett:2000ha} and in the case of single-field tachyonic
preheating in \Ref{Suyama:2006sr}).

These phenomena can lead to radical shifts in the standard picture of
how reheating proceeds. Indeed, in \Ref{Martin:2019nuw}, it was shown
that the production of light PBHs from metric preheating is so
efficient that they can quickly come to dominate the universe content,
such that reheating no longer occurs because of the inflaton decay, as
previously described, but rather through PBHs Hawking
evaporation. This conclusion, however, was reached by neglecting the
decay products of the inflaton throughout the instability phase, and
by simply assuming that they would terminate the instability abruptly
at the time when they dominate the energy budget (if PBHs have not
come to dominate the universe before then). However, as will be made
explicit below, the instability of metric preheating proceeds in the
narrow resonance regime. One may therefore be concerned that it
requires a delicate balance in the dynamics of the system, and that
even a small amount of produced radiation could be enough to distort
or jeopardise the instability mechanism. The goal of this paper is
therefore to investigate how the presence of inflaton decay products
(modelled as a perfect fluid), produced by perturbative reheating,
affects the metric preheating instability.

The paper is organised as follows. In \Sec{sec:reheating}, we briefly
review metric preheating, which leads to the growth of the inflaton
density contrast at small scales. Then, in \Sec{sec:radiativedecay},
we study whether a small amount of radiation, originating from the
inflaton decay, can modify this growth. For this purpose, we introduce
a covariant coupling model between the inflaton scalar field and a
perfect fluid, leading to equations of motion at the background (see
\Sec{subsec:back}) and perturbative (see \Sec{subsec:pert}) levels
that feature no substantial change in the instability structure until
the fluid dominates. In \Sec{subsec:pbh}, we discuss the application of
the previous results to the production of PBHs during reheating,
which, we stress, cannot be described as originating from perfect
fluid inhomogeneities, contrary to what is sometimes argued. Finally,
in \Sec{sec:conclusions}, we briefly summarise our main results and
present our conclusions.
\section{Preheating in single-field inflation}
\label{sec:reheating}
In this work, we consider single scalar field models of inflation,
with a canonical kinetic term. In these models, a homogeneous inflaton
field $\phi(t)$ drives the expansion of a flat
Friedmann-Lema\^itre-Robertson-Walker (FLRW) space-time, described by
the metric $\dd s^2=-\dd t^2+a^2(t)\dd {\bm x}^2$, where $a(t)$ is the
FLRW scale factor. The corresponding equations of motion are the
Friedmann and Klein-Gordon equations, namely
\begin{align}
\label{eq:KG&F}
H^2=\frac{1}{3\Mp^2}\left[\frac{\dot{\phi}^2}{2}+V(\phi)\right], \quad
\ddot{\phi}+3H\dot{\phi}+V_\phi\left(\phi\right) =0\, ,
\end{align}
where $H=\dot{a}/a$ is the Hubble parameter, $V_\phi$ the derivative
of the potential with respect to $\phi$, $\Mp$ the reduced Planck mass
and a dot denotes a derivative with respect to cosmic time $t$. The
inflaton field potential $V(\phi)$ must be such that the potential
energy dominates over the kinetic energy of the inflaton, and
inflation ($\ddot{a}>0$) ends when they become comparable, that is to
say when the first slow-roll parameter $\epsilon_1 \equiv
-\dot{H}/H^2$ reaches one. This usually happens in the vicinity of a
local minimum of the potential. There, most potentials can be
approximated by a quadratic function, $V(\phi)\sim m^2\phi^2/2$, where
$m$ is the curvature of the potential at its minimum. In fact, this
expression can be seen as a leading-order Taylor expansion of the
potential around its minimum, and it is not valid only for potentials
having an exactly vanishing mass at their minimum, for which the
leading term is of higher order. When the inflaton reaches this region
of the potential, it oscillates according to $\phi(t)\propto a^{-3/2}
\sin\left(mt\right)$, the expansion becomes, on average, decelerated,
and similar to that of a matter-dominated
universe~\cite{Turner:1983he}, \ie $\langle \rho \rangle \propto
a^{-3}$ (where $\langle \cdot \rangle$ denotes averaging over one
oscillation).
\subsection{Perturbative reheating}
\label{sec:perturbative:reheating}
These considerations however ignore the possible coupling of the
inflaton with other degrees of freedom. In order to incorporate it,
several descriptions are possible. A simple way, which corresponds to
``perturbative reheating'', consists in introducing a term ``$\Gamma
\dot{\phi}$'' (where $\Gamma$ is a decay rate) in the Klein-Gordon
equation to account for the decay of the inflaton into a perfect fluid
(typically
radiation)~\cite{Albrecht:1982mp,Dolgov:1982th,Abbott:1982hn,Kofman:1997yn}.
In this case, the friction term becomes
$(3H+\Gamma/2)\dot{\phi}$. Initially, $H\gg \Gamma$ and the effect of
the inflaton decay is negligible, until $H$ crosses down $\Gamma$, at
a time around which most of the decay of the inflaton occurs. In the
next section, we explain how to introduce $\Gamma$ covariantly, thus
allowing us to perform a consistent treatment both at the background
and perturbative levels. Microscopically, if one considers for
instance that $\phi$ is coupled to another scalar field $\chi $
through the interaction Lagrangian ${\cal L}_\mathrm{int}=-2g^2\sigma
\phi \chi^2$, where $g$ is a dimensionless coupling constant and
$\sigma$ a new mass scale, the corresponding decay rate can be
calculated within perturbation theory and one finds
$\Gamma=g^4\sigma^2/(4\pi m)$~\cite{Kofman:1997yn}.  If this process
occurs at sufficiently high energy, the mass of the $\chi$-particles
are small compared to the Hubble parameter at decay and, effectively,
the inflaton field decays into relativistic matter or radiation.

\subsection{Non-perturbative preheating}
\label{sec:NonPerturbativePreheating}

\begin{figure}[t]
\begin{center}
\includegraphics[width=0.6\textwidth]{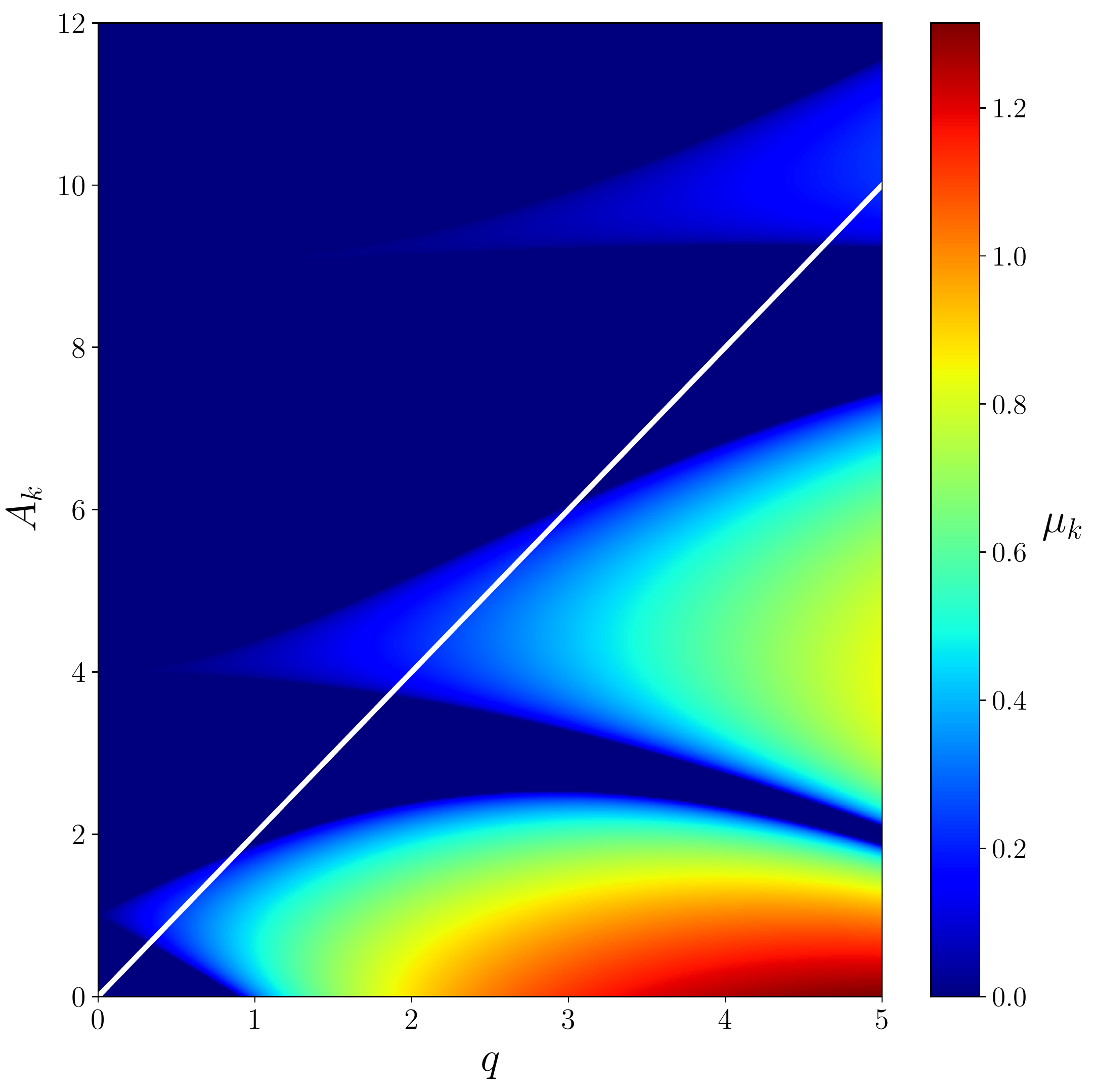}
\caption{Instability chart of the Mathieu equation. The colour code
  (see the colour bar on the right hand side of the plot) represents
  the value of the Floquet exponent $\mu_{\bm k}$ of the unstable
  mode. In the present case, stable solutions corresponds to $\mu_{\bm
    k}=0$ and are represented by the dark blue regions. The other
  regions, structured in different bands, correspond to unstable
  solutions.}
\label{fig:mapmathieu}
\end{center}
\end{figure}

The above perturbative description is however not sufficient since
non-perturbative effects can also play an important
role~\cite{Shtanov:1994ce, Kofman:1994rk, Kofman:1997yn}. This can be
simply illustrated if one considers the case where the interaction
Lagrangian reads ${\cal L}_\mathrm{int}=-g^2\phi^2\chi^2/2$. If one
denotes the monotonously decreasing amplitude of the inflaton
oscillations as $\phi_0(t)$, such that $\phi\simeq\phi_0(t)\sin(mt)$,
then the equation of motion of the Fourier transform $\chi_{\bm k}$ of
the field $\chi$ reads
\begin{align}
\label{eq:eom:chik:1}
\ddot{\chi}_{\bm k}+ 3 H \dot{\chi}_{\bm k}+\left[
  \frac{k^2}{a^2(t)}+m_\chi^2+ g^2\phi_0^2(t)\sin^2(m
  t)\right]\chi_{\bm k}=0\, ,
\end{align}
where $m_\chi$ is the mass of $\chi$ and ${\bm k}$ the wavenumber of
the mode under consideration. Writing $X_{\bm k}=a^{3/2}\chi_{\bm k}$
and using the variable $z\equiv mt$, the above equation can also be
written under the following form
\begin{align}
  \label{eq:mathieu}
  \frac{\dd ^2 X_{\bm k}}{\dd z^2}
  +\left[A_{\bm k}-2q\cos\left(2z\right)\right]
  X_{\bm k}=0,
  \end{align}
where the quantities $A_{\bm k}$ and $q$ are defined by
\begin{align}
\label{eq:A:q:def}
A_{\bm k} &=\frac{k^2}{a^2m^2}+\frac{m_\chi^2}{m^2}
-\frac32 \frac{H^2}{m^2}\left(\frac32-\epsilon_1\right)
+2q, \quad
  q=\frac{g^2\phi_0^2}{4 m^2}.
  \end{align}
As a first step, in order to gain intuition about the behaviour of the
solutions, it is convenient to analyse the above equation in the
Minkowski space-time (for simplicity, we also consider the massless
case $m_\chi=0$). In that situation, the coefficients $A_{\bm
  k}=k^2/m^2+2q$ and $q$ are constant and \Eq{eq:mathieu} is a Mathieu
equation~\cite{Abramovitz:1970aa:Mathieu}. This equation possesses
unstable, exponentially growing solutions $\chi_{\bm {k}} \propto
\exp(\mu_{\bm{k}} z)$. In \Fig{fig:mapmathieu}, known as the Mathieu
instability chart, we display the value of $\mu_{\bm{k}}$, the
so-called Floquet index of the unstable mode (namely the maximum of
the two Floquet indices), as a function of $A_{\bm k}$ and
$q$. Unstable regions correspond to where $\mu_{\bm k}>0$, and are
organised in several ``bands'', which can be identified as the non
dark-blue regions in \Fig{fig:mapmathieu}. Since $A_{\bm
  k}=k^2/m^2+2q$, the parameter space of interest is such that $A_{\bm
  k}> 2q$, which corresponds to the region above the white line in
\Fig{fig:mapmathieu}. At a given $q$, one can see in
\Fig{fig:mapmathieu} that there are several ranges of values of
$A_{\bm k}$, hence several ranges of wavenumbers $k$, where an
instability develops. One also notices that the band with the smallest
value of $A_{\bm k}$ is the most pronounced one. When $q\gg 1$, the
range of excited modes is large, which corresponds to being in the
``broad-resonance'' regime. When $q\ll 1$, on the contrary, there is
only a small range of values of $k$ being excited, which correspond to
the ``narrow-resonance'' regime. In that limit, the boundaries of the
first band correspond to $1-q\lesssim A_{\bm k}\lesssim 1+q$.

Then, space-time dynamics must be restored and its impact on the
previous considerations discussed. In that case, three time scales
play a role in \Eq{eq:eom:chik:1}: the inflaton oscillation period
$m^{-1}$, the Hubble time $H^{-1}$, and the $\bm{k}$-mode period,
$a/k$. The quantities $A_{\bm k}$ and $q$ now become functions of time
[notice that the oscillating phase starts when $m\sim H$, and since
  $H$ decreases afterwards, one quickly reaches the regime where $H\ll
  m$ and, as a consequence, the term $\propto H^2/m^2$ in the
  definition~(\ref{eq:A:q:def}) of $A_{\bm k}$ can be neglected]. This
means that \Eq{eq:mathieu} is no longer a Mathieu equation: a given
mode $\bm{k}$ now follows a certain path in the map of
\Fig{fig:mapmathieu}. What is then the fate of the two regimes (narrow
and broad resonance) identified before? Since more time is being spent
in the wide bands than in the narrow ones, the broad resonance regime
is the most important one to amplify the $\chi$ field. However, this
regime is also crucially modified by space-time expansion and gives
rise to the so-called ``stochastic-resonance regime'', discovered in
\Ref{Kofman:1997yn}. Preheating effects have also been studied in
other contexts, for instance when the curvature of (some region of)
the inflationary potential is negative, as it is the case, for
instance, in small-field inflation, leading to tachyonic
preheating~\cite{Felder:2000hj, Desroche:2005yt}.\\

\subsection{Metric preheating}
So far we have discussed preheating at the background level only,
without including the inflaton and metric perturbations. They however
play an important role, in a mechanism known as ``metric
preheating''~\cite{Finelli:1998bu,Bassett:1999cg,Bassett:1999mt,
  Jedamzik:1999um,Finelli:2000ya,Allahverdi:2010xz}. Including scalar
fluctuations only, in the longitudinal gauge, the perturbed metric can
be written as $\dd s^2=a^2(\eta)\left[-\left(1+2\Phi\right) \dd
  \eta^2+\left(1-2\Phi\right)\delta _{ij}\dd x^i\dd x^j\right]$, where
$\eta $ is the conformal time related to the cosmic time by $\dd
t=a\dd \eta$. As is apparent in the previous expression, the scalar
perturbations are described by a single quantity, namely the Bardeen
potential $\Phi$. Matter perturbations, which, in the context of
inflation, boil down to scalar field perturbations, are also
characterised by a single quantity, the perturbed scalar field $\delta
\phi^{\mathrm{(gi)}}$, where the ``gi'' indicates that this is a
gauge-invariant quantity ($\delta \phi^{\mathrm{(gi)}}=\delta\phi$ in
the longitudinal gauge and is mapped by gauge transformations
otherwise). Using the perturbed Einstein equations, the whole scalar
sector can in fact be described by a single quantity, which is a
combination of metric and matter perturbations. This single quantity
is the Mukhanov-Sasaki variable~\cite{Mukhanov:1981xt,Kodama:1985bj}
$v\equiv a\left[\delta \phi^{\mathrm{(gi)}}+\phi'\Phi/{\cal
  H}\right]$, where ${\cal H}=a'/a$ (a prime denotes a derivative with
respect to conformal time) is the conformal Hubble parameter, and is
directly related to the comoving curvature perturbation $\mathcal{R}$
by $v=Z\mathcal{R}$, where $Z\equiv \sqrt{2\epsilon_1}a\Mp$. The
Fourier component $v_{\bm k}$ evolves according to the equation of a
parametric oscillator where the time dependence of the frequency is
determined by the dynamics of the background~\cite{Mukhanov:1990me}
\begin{align}
  \label{eq:eomv}
  v_{\bm k}''+\left(k^2-\frac{Z''}{Z}\right)v_{\bm k} = 0,
\end{align}
with
\begin{align}
\label{eq:MS}
\frac{Z''}{Z}= a^2 H^2 \left[ \left(1+\frac{\epsilon_2}{2}\right)
  \left(2-\epsilon_1+\frac{\epsilon_2}{2}\right)
  +\frac{\epsilon_2\epsilon_3}{2}\right]\, ,
\end{align}
where $\epsilon_2 \equiv \dd\ln\epsilon_1/\dd N$ and $\epsilon_3
\equiv \dd\ln\epsilon_2/\dd N$ are the second and the third slow-roll
parameters respectively.

The question is then whether \Eq{eq:eomv} allows for parametric
resonance when the inflaton field oscillates at the bottom of its
potential. One might indeed expect that the oscillations in $\phi(t)$
induce oscillations in the Hubble parameter $H$, hence in the
slow-roll parameters, hence in $Z''/Z$. In this case, \Eq{eq:eomv}
could be of the Mathieu type, or more generally of the Hill type, and
could lead to parametric resonance. This was first thought not to be
the case, the main argument being that, despite the oscillations in
$Z''/Z$, the curvature perturbation has to remain constant and, as a
consequence, there cannot be any growth of scalar
perturbations~\cite{Finelli:1998bu}. It has also been stressed that
the situation can be drastically different in multiple-field
inflation~\cite{Finelli:2000ya}, where entropy fluctuations source the
evolution of curvature perturbations. If the entropy fluctuations are
parametrically amplified, they can also cause a parametric
amplification of adiabatic perturbations. This is the reason why
metric preheating was first mostly studied in the context of
multiple-field (and in practice, mostly two-field) inflation, see for
instance \Ref{Finelli:2000ya}.

It was then realised in \Ref{Jedamzik:2010dq} (see also
\Ref{Easther:2010mr}) that \Eq{eq:eomv} can be put under the form
\begin{align}
\label{eq:Mathieu:v}
  \frac{\dd^2}{\dd z^2}\left(\sqrt{a} v_{\bm k}\right)
  +\left[A_{\bm k}-2q\cos(2z)\right]\left(\sqrt{a} v_{\bm k}\right)=0,
\end{align}
with
\begin{align}
\label{eq:A:q:def:metric:preheating}
  A_{\bm k}=1+\frac{k^2}{m^2a^2}, \quad
  q=\frac{\sqrt{6}}{2}\frac{\phi_\uend}{\Mp}\left(\frac{a_\uend}{a}\right)^{3/2},
\end{align}
where $a_\uend$ is the scale factor at the end of inflation and
$z\equiv mt+\pi/4$. Although, strictly speaking, this equation is not
of the Mathieu type because of the time dependence of the parameters
$A_{\bm k}$ and $q$, it was shown in \Ref{Jedamzik:2010dq} that this
time dependence is sufficiently slow so that \Eq{eq:Mathieu:v} can be
analysed using Mathieu equations techniques. At the end of inflation
and at the onset of the oscillations,
$\phi_0(t_\mathrm{end})=\phi_\mathrm{end}$ is of the order of the
Planck mass, so \Eq{eq:A:q:def:metric:preheating} indicates that $q$
starts out being of order one and quickly decreases afterwards. In
contrast to the situation of non-perturbative preheating discussed in
\Sec{sec:NonPerturbativePreheating}, the narrow-resonance regime $q\ll
1$ is therefore always the relevant one for metric preheating. In that
regime, and contrary to the case of broad resonance, space-time
expansion does not blur the resonance but, on the contrary, reinforces
its effectiveness, in a sense that we will explain below. As mentioned
above, in the $q\ll 1$ limit, the boundaries of the first instability
band are given by $1-q < A_{\bm k} < 1+q$, which here translates into
\begin{align}
\label{eq:instability:band:1}
k < a \sqrt{3 H m}.
\end{align}
One notices the appearance of a new scale in the problem, namely
$\sqrt{3Hm}$. Since the universe behaves as matter dominated during
the oscillations of the inflaton, $a\sqrt{H} \propto a^{1/4}$, and the
upper bound~\eqref{eq:instability:band:1} increases with time. This
means that the range of modes subject to the instability widens up as
time proceeds, hence the above statement that space-time expansion
strengthens the resonance effect.

\begin{figure}[t]
\begin{center}
  \includegraphics[width=1.0\textwidth]{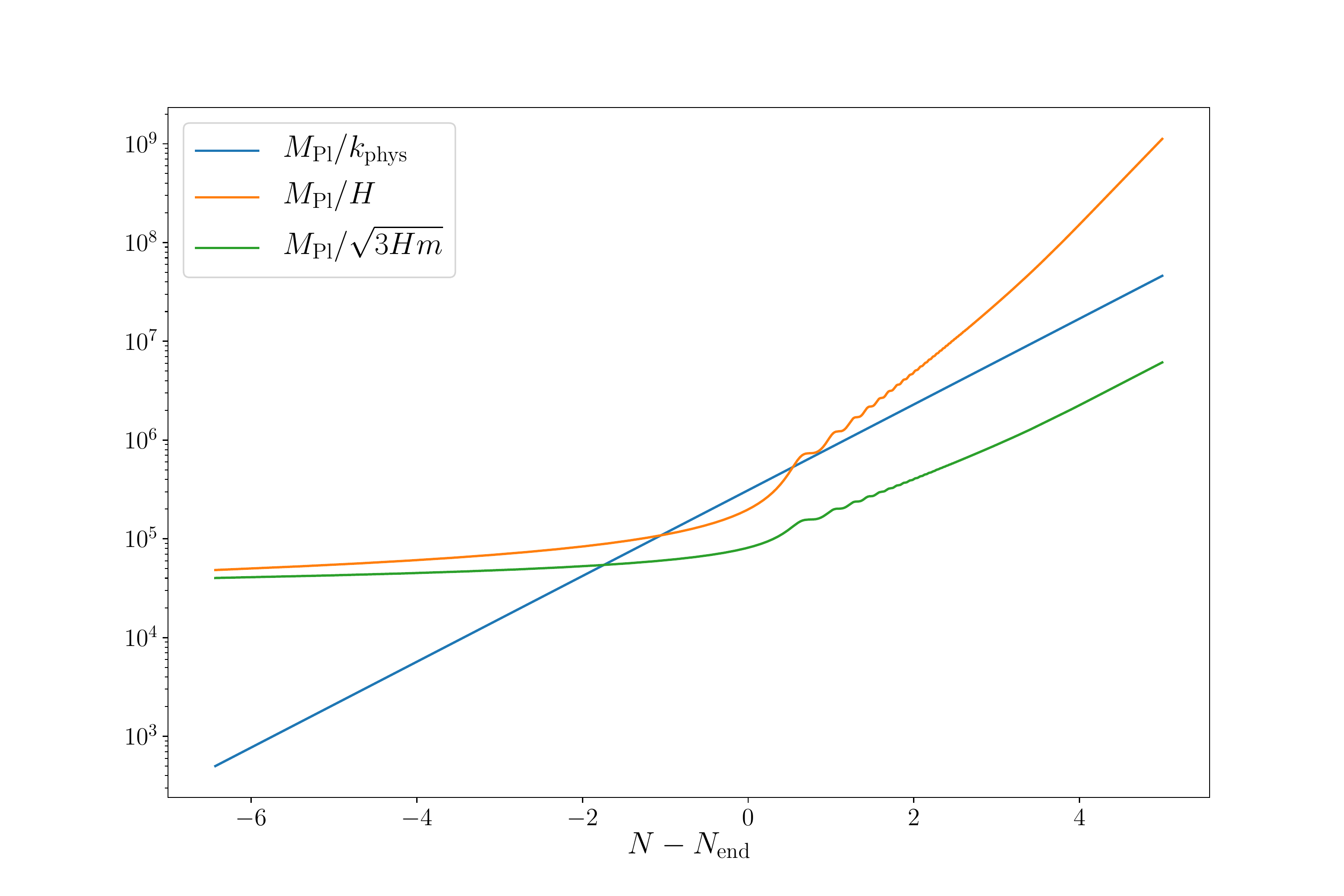}
\caption{Evolution of the physical scales appearing in
  \Eq{eq:instability:band:2}, with time parameterised by the number of
  \efolds~(counted from the end of inflation). The orange solid line
  represents the Hubble radius $1/H$, the solid green line the new
  length scale $1/\sqrt{3Hm}$ and the solid blue line the physical
  wavelength of a mode of interest, which enters the instability band
  from above. In all figures of this work, we study the comoving scale
  $k/a_\mathrm{ini}=0.002 \Mp$, where the initial time of integration
  is set 6 \efolds~before the end of inflation, in a quadratic
  potential model $V(\phi)=m^2\phi^2/2$ with $m=10^{-5}\Mp$. The
  inflaton decay constant (the definition of which is detailed in
  \Sec{subsec:back}) is given by $\Gamma=10^{-7}\Mp$.  Here, we
  consider the case where the inflaton decays into a radiation fluid,
  with equation-of-state parameter $w_{\mathrm{f}}=1/3$.}
\label{fig:scale}
\end{center}
\end{figure}
Inside the first instability band, the Floquet index of the unstable
mode is given by $\mu_{\bm{k}} \simeq q/2$, so $v_{\bm{k}}\propto
a^{-1/2} \exp(\int \mu_{\bm{k}} \dd z)\propto
a$~\cite{Finelli:1998bu,Jedamzik:2010dq}. The comoving curvature
perturbation, $\mathcal{R}_{\bm k}= v_{\bm k}/(\Mp a \sqrt{2
  \epsilon_1})$, is thus conserved for modes satisfying
\Eq{eq:instability:band:1}. Notice that, since $H\ll m$ during the
oscillatory phase, this comprises super-Hubble modes, $k<a H$, for
which the conservation of $\mathcal{R}$ is a well-known
result~\cite{Mukhanov:1981xt, Kodama:1985bj}. However, the
conservation of $\mathcal{R}$ also applies for those sub-Hubble modes
having $k < a \sqrt{3 H m}$, and for which this leads to an increase
of the density contrast. Indeed, if $\mathcal{R}$ is constant, and
given that the pressure vanishes on average, the fractional energy
density perturbation $\delta_{\bm k} = \delta\rho_{\bm k}/\rho$ (where
$\rho$ is the background energy density of the scalar field) in the
Newtonian gauge is related to the curvature perturbation
via~\cite{Jedamzik:2010dq}
\begin{align}
  \delta_{\bm k} = -\frac{2}{5}
  \left(\frac{k^2}{a^2 H^2}+3\right) \mathcal{R}_{\bm{k}}\, .
\end{align}
On super-Hubble scales, the first term in the braces can be neglected,
hence $\delta_{\bm{k}}$ is constant as $\mathcal{R}_{\bm {k}}$. On
sub-Hubble scales however, the first term becomes the dominant one,
and since $a^2 H^2\propto a^{-1}$, the density contrast grows like
\begin{align}
\label{eq:delta:a}
\delta_{\bm{k}}\propto a\, .
\end{align}
This corresponds to a physical instability (notice that, at sub-Hubble
scales, there are no gauge ambiguities in the definition of the
density contrast), which therefore operates at scales
\begin{align}
\label{eq:instability:band:2}
aH<k<a\sqrt{3 H m}.
\end{align}
The scales appearing in this relation are displayed in
\Fig{fig:scale}. An instability is triggered if the physical
wavelength of a mode (blue line) is smaller than the Hubble radius
(orange line) during the oscillatory phase and larger than the new
scale $1/\sqrt{3Hm}$ (green line). This implies that the instability
only concerns modes that are inside the Hubble radius at the end of
the oscillatory phase, which is not the case for the scales probed in
the CMB. It is therefore true that metric preheating does not operate
at CMB scales, although it plays a crucial role at smaller scales
(typically those crossing out the Hubble radius a few \efolds~before
the end of inflation) where it triggers an instability in the
narrow-resonance regime. The growth of the density contrast along
\Eq{eq:delta:a} may have several important consequences such as early
structure formation~\cite{Jedamzik:2010dq}, emission of gravitational
waves~\cite{Jedamzik:2010hq} and, as recently studied in
\Ref{Martin:2019nuw}, formation of PBHs that may themselves contribute
to the reheating process, via Hawking evaporation.

As already mentioned, preheating effects cannot by themselves ensure a
complete transition to the hot big-bang phase~\cite{Kofman:1997yn,
  Bassett:2005xm,Lozanov:2017hjm,Maity:2018qhi} (except if reheating
occurs by Hawking evaporation of the very light primordial black holes
produced from the instability if they come to dominate the universe
content~\cite{Martin:2019nuw}), which also requires perturbative decay
of the inflaton to complete. Metric preheating has however been
investigated only in the context of purely single-field setups, and it
is not clear whether or not the narrow resonant structure of metric
preheating is immune to the decay of the inflaton into other degrees
of freedom. This is why in the next sections, we study metric
preheating and perturbative reheating altogether, in order to
determine if and how the later can spoil the former.
\section{Metric preheating and radiative decay}
\label{sec:radiativedecay}
We have seen before that perturbations entering the instability
band~\eqref{eq:instability:band:2} undergo a growth of their density
contrast proportional to the FLRW scale factor, see \Eq{eq:delta:a},
and that this can lead to a variety of interesting phenomena. At some
stage, however, the inflaton field decays and the growth of the
density contrast, sourced by the oscillations of the inflaton
condensate, should come to an end. In \Ref{Martin:2019nuw} this was
simply modelled by abruptly stopping the oscillating phase at a
certain time (e.g., when $H$ becomes smaller than a certain value that
can be identified with the decay rate $\Gamma$) and by assuming
instantaneous production of radiation at that time. However, clearly,
the inflaton decay should proceed continuously. Although it is true
that the production of radiation becomes sizeable when the Hubble
parameter becomes of the order of the decay rate, tiny amounts of
radiation are present before and one may wonder whether or not they
can destroy the delicate balance which is responsible for the presence
of the modes in the instability band. Indeed, the instability proceeds
in the narrow resonance regime, which means that the instability band
spans a small, fine-tuned volume of parameter space. In this section,
we investigate these questions.
\subsection{Setup and background}
\label{subsec:back}
In order to study the influence of fluid production, we must first
modify the equations of motion of the system and introduce an explicit
coupling between the inflaton field and a perfect fluid, both at the
background and perturbative levels. This poses non-trivial problems at
the technical level and we now review the formalism that can be used
in order to tackle them. Let us consider a collection of fluids in
interaction. The presence of interactions break the energy-momentum
conservation for each fluid. On very general grounds, their
non-conservations can be described non-perturbatively by detailed
balance equations of the form \cite{Malik:2004tf, Choi:2008et,
  Malik_2009, Leung:2012ve, Leung:2013rza, Huston:2013kgl,
  Visinelli:2014qla}
\begin{align}
  \label{eq:conservationT}
\nabla _{\nu}T^{\mu \nu}_{(\alpha)}=\sum_{\beta}\left[Q^{\mu}_{(\alpha)\rightarrow (\beta)}
-Q^{\mu}_{(\beta)\rightarrow (\alpha)}\right],
\end{align}
where the transfer coefficients $Q^{\mu}_{(\alpha)\rightarrow
  (\beta)}$ are responsible for the non-conservation of energy-momentum
originating from the interaction between the fluids. The indices
between parenthesis [such as ``$(\alpha)$''] label the different fluid
components. The term $Q^{\mu}_{(\alpha)\rightarrow (\beta)}$ describes
a loss due to the decay of the fluid $\alpha$ into the fluids $\beta$
while, on the contrary, the term $Q^{\mu}_{(\beta)\rightarrow
  (\alpha)}$ corresponds to a gain originating from the decay of the
fluids $\beta$ into $\alpha$. The evolution of the stress-energy
tensor of the fluid $\alpha$, which is denoted $T^{\mu
  \nu}_{(\alpha)}$, is then controlled by the detailed balance between
those two effects. The transfer coefficient
$Q^{\mu}_{(\alpha)\rightarrow (\beta)}$ can always be decomposed as
\begin{align}
\label{eq:defQ}
Q^{\mu}_{(\alpha)\rightarrow (\beta)}=Q_{(\alpha)\rightarrow (\beta)}u^{\mu}
+f^{\mu}_{(\alpha)\rightarrow (\beta)},
\end{align}
where $Q_{(\alpha)\rightarrow (\beta)}$ is a scalar quantity and
$f^{\mu}_{(\alpha)\rightarrow (\beta)}$ a vector orthogonal to the
matter flow, that is to say $f^{\mu}_{(\alpha)\rightarrow
  (\beta)}u_{\mu}=0$ where $u^{\mu}$ is the total velocity of matter.
In an FLRW universe it is given by $u^{\mu}=(1/a,{\bm 0})$,
$u_\mu=(-a,{\bm 0})$, which immediately implies that
$f^0_{(\alpha)\rightarrow (\beta)}=0$ at the background
level. Furthermore, in an homogeneous and isotropic background, one
must have $f^i_{(\alpha)\rightarrow (\beta)}=0$ to respect the
symmetries of space-time, hence $f^{\mu}_{(\alpha)\rightarrow
  (\beta)}=0$. This allows us to write $Q^0_{(\alpha)\rightarrow
  (\beta)}=Q_{(\alpha)\rightarrow (\beta)}/a$ and
$Q^i_{(\alpha)\rightarrow (\beta)}=0$. At the background level, the
energy transfer is therefore entirely specified by the scalar
$Q_{(\alpha)\rightarrow (\beta)}$.

Let us now apply these considerations to a system made of one scalar
field (the inflaton field) and a perfect fluid assumed to be the
inflaton decay product. In order to consistently couple the scalar
field $\phi$ with the fluid, one must view the scalar field as a
collection of two fictitious fluids, the ``kinetic'' one, with energy
density and pressure given by $\rho_K=p_K=\phi'^2/(2a^2)$, and the
``potential'' one, with $\rho_V=-p_V=V(\phi)$, each of them having a
constant equation-of-state, one and minus one, respectively. The
energy density and pressure of the scalar field are just the sums of
the energy densities and pressures of the two fluids, namely
$\rho_\phi=\rho_K+\rho_V$ and $p_\phi=p_K+p_V$. In order to recover
the standard equations for a scalar field, one must also consider that
the fictitious kinetic and potential fluids are coupled, the coupling
being described by~\cite{Malik:2004tf}
\begin{align}
aQ_{K\rightarrow V}&=-\phi'V_{\phi}, \quad aQ_{V\rightarrow
  K}=0.
\end{align}
\begin{figure}[t]
\begin{center}
  \includegraphics[width=1.0\textwidth]{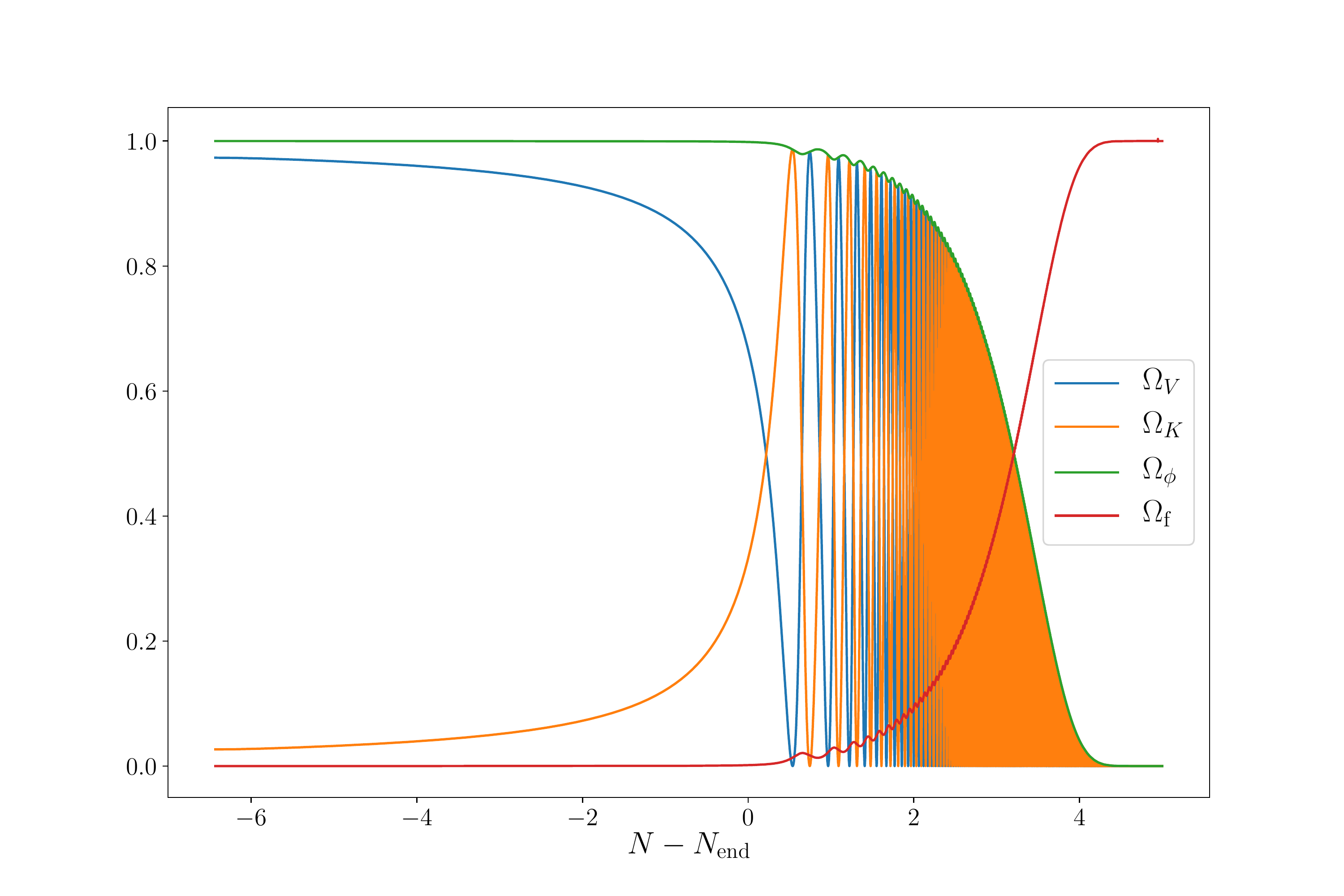}
  \caption{Evolution of the different energy density contributions as
    a function of the number of \efolds. The solid orange line
    represents the contribution of the fictitious kinetic fluid, the
    solid blue line the contribution of the fictitious potential fluid
    and the solid green line the contribution of the physical scalar
    field which is the sum of those two. The solid red line
    corresponds to the contribution of radiation. Before the end of
    inflation, the scalar field dominates the energy budget and, then,
    when its decay becomes effective, radiation takes over. The
    parameter values are the same as in \Fig{fig:scale}.}
\label{fig:backgroundOmega}
\end{center}
\end{figure}

Then, we consider the ``real'' interaction between the scalar field
and the perfect fluid (in practice radiation). The crucial
idea~\cite{Malik:2004tf, Jerome_Lucas_in_prep} is that it is obtained by
coupling the fluid only to the fictitious kinetic fluid related to
$\phi$ and introduced above (and not to the potential fluid). This
implies that $Q^{\mu}_{V\rightarrow
  \mathrm{f}}=Q^{\mu}_{\mathrm{f}\rightarrow V}=0$. In practice, we
consider the case where the covariant interaction between ``$K$'' and
``f'' can be described non-perturbatively by the following energy
four-momentum transfer:
\begin{align}
\label{eq:QmuK:to:f}
Q^{\mu}_{K \rightarrow \mathrm{f}}=\Gamma
\, T^{\mu \nu}_{K}u_{\nu}^{K}, \quad Q^{\mu}_{\mathrm{f}\rightarrow K}=0,
\end{align}
where $\Gamma$ is the decay rate and is the only new parameter
introduced in order to account for the interaction. Note that this
  description should be understood as a phenomenological
  parametrisation of the decays of scalar fields in cosmological
  fluids, and not as a concrete microphysical model. At the
background level, one recovers the picture described in
\Sec{sec:perturbative:reheating}, since the equations of
motion~\eqref{eq:conservationT} of the system (namely the energy
conservation equation, since the momentum conservation equation is
trivial) can be written as
\begin{align}
  \label{eq:eomscalargamma}
  \phi''+2{\cal H}\phi'+\frac{a\Gamma}{2}\phi'+a^2V_\phi&=0, \\
\label{eq:eomfluidgamma}
  \rho_\mathrm{f}'+3{\cal H}(1+w_\mathrm{f})\rho_\mathrm{f} -\frac{\Gamma}{2a}\phi'^2&=0.
  \end{align}
The first equation is the modified Klein-Gordon equation while the
second one is the modified conservation equation for the fluid with
equation-of-state parameter $w_\mathrm{f}$ (in practical applications,
unless stated otherwise, we take $w_\mathrm{f}=1/3$). These equations
are usually introduced in a phenomenological way. The fact that we are
able to derive them from a covariant formulation,
\Eq{eq:conservationT}, will allow us to describe perturbations in the
same framework, by assuming that \Eq{eq:conservationT} also holds at
the perturbative (and in principle, even non-perturbative) level, see
\Sec{subsec:pert}.

\begin{figure}[t]
\begin{center}
  \includegraphics[width=1.0\textwidth]{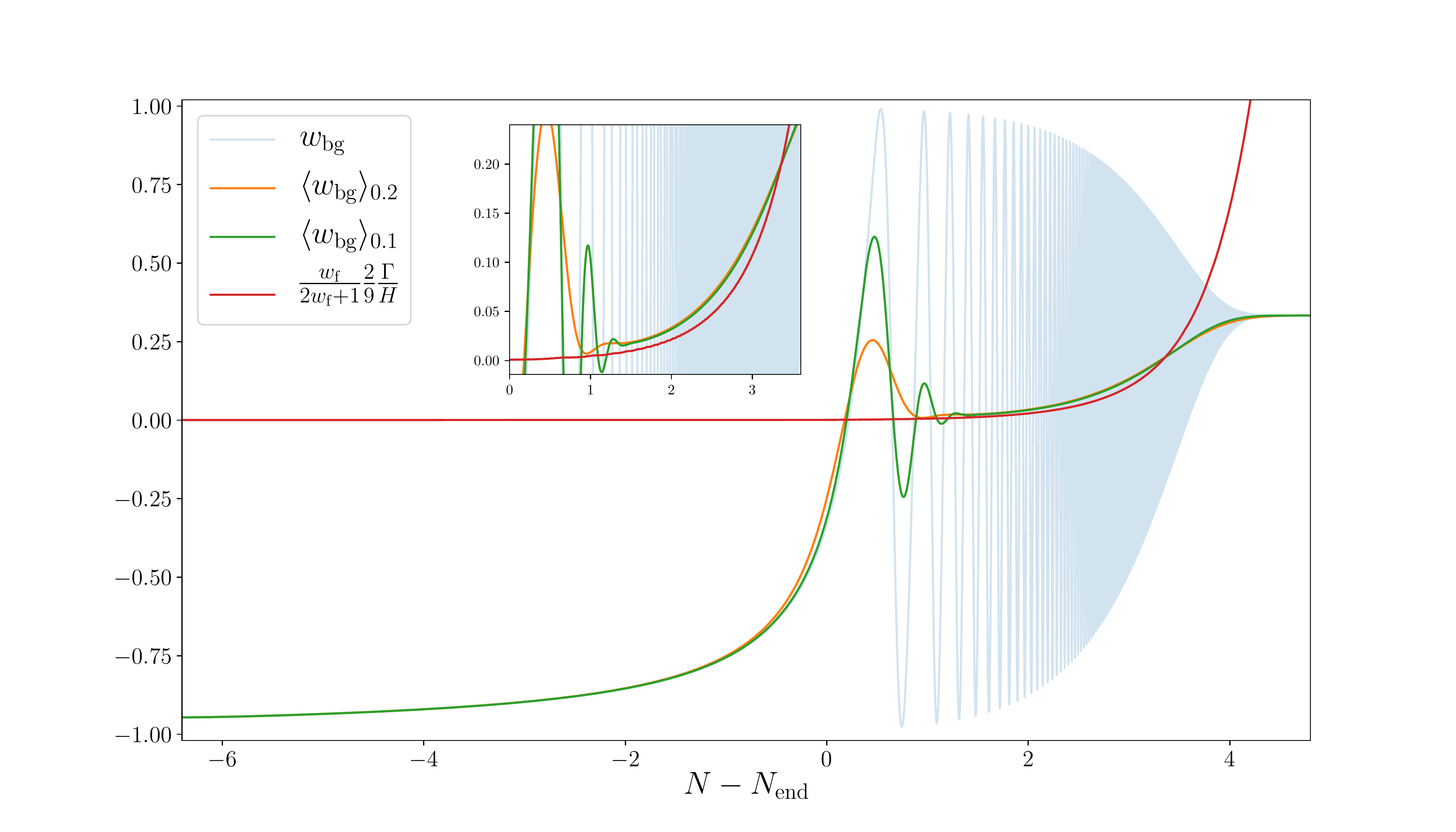}
  \caption{Evolution of the instantaneous (transparent blue line) and
    time-averaged equation-of-state parameters as a function of the
    number of \efolds, in the same setup as in
    \Fig{fig:backgroundOmega}. The averaging procedure consists in
    convolving the instantaneous signal with a Gaussian kernel of
    constant standard deviation given by $0.2$ \efolds~(orange line)
    and $0.1$ \efolds~(green line), such that oscillations on shorter
    time scales are averaged out. The analytical approximation
    \Eq{eq:eos_anal} is also displayed (red line), and the inset plot
    zooms in its regime of validity (\ie at the onset of the
    oscillating phase).}
\label{fig:backgroundw}
\end{center}
\end{figure}

We have numerically integrated \Eqs{eq:eomscalargamma}
and~\eqref{eq:eomfluidgamma} for $V(\phi)=m^2\phi^2/2$ with
$m=10^{-5}\Mp$, $w_\mathrm{f}=1/3$ and $\Gamma=10^{-7}\Mp$. For a
quadratic potential, inflation stops when $\phi_\uend/\Mp\simeq
\sqrt{2}$ and the (slow-roll) trajectory reads $\phi(N)/\Mp\simeq
\sqrt{2-4(N-N_\uend)}$ where $N$ is the number of \efolds. Here, we
want to focus on the last \efolds~of inflation and, therefore, the
initial conditions are chosen such that the evolution is started on
the slow-roll attractor at $\phi_\uini/\Mp\simeq 5$, corresponding to
$N_\uend-N_\uini\simeq 6$, and $\rho_\mathrm{f}^\uini=0$ (as we will
show below, the precise choice of the time at which we set
$\rho_\mathrm{f}=0$ does not matter since $\rho_\mathrm{f}$ quickly
reaches an attractor during inflation). The result is represented in
\Figs{fig:backgroundOmega} and~\ref{fig:backgroundw}, where inflation
ends when $N-N_\uend=0$. Then starts the oscillation phase. In
\Fig{fig:backgroundOmega}, $\Omega_K\equiv
\rho_K/(\rho_\phi+\rho_\mathrm{f})$, $\Omega_V\equiv
\rho_V/(\rho_\phi+\rho_\mathrm{f})$, $\Omega _\phi\equiv
\Omega_K+\Omega_V$ and $\Omega_\mathrm{f}\equiv
\rho_\mathrm{f}/(\rho_\phi+\rho_\mathrm{f})$ are displayed as a
function of time. Initially, we have $\Omega_\phi\simeq \Omega_V\simeq
1$ and $\Omega_\mathrm{f}\simeq\Omega_K\simeq 0$. Indeed, in the
slow-roll phase, the potential energy largely dominates over the
kinetic energy, since the first slow-roll parameter can be expressed
as $\epsilon_1 = 3[(1+w_\mathrm{f})\Omega_\mathrm{f}+2\Omega_K]/2$,
hence both $\Omega_\mathrm{f}$ and $\Omega_K$ need to be very
small. In this regime, we also have $H\gg \Gamma$ and the amount of
radiation being produced is very small. Then, inflation stops and
$\Omega_K$ and $\Omega_V$ become of comparable magnitude and start
oscillating. During that phase, radiation still provides a small,
though non-vanishing, contribution. Finally, when $H\simeq \Gamma$, at
the time $N\equiv N_\Gamma$, radiation starts to be produced in a
sizeable amount and cannot be neglected anymore. After the end of
inflation, the universe expands, on average, as in a matter-dominated
era, for which $H\propto a^{-3/2}$, that is to say $H\propto H_\uend
\exp[-3(N-N_\uend)/2]$. Writing the condition $H=\Gamma$ thus leads to
an estimate of $N_\Gamma$, namely
\begin{align}
  N_\Gamma- N_\uend \simeq \frac{2}{3}\ln \left(\frac{\sqrt{2}}{2}
  \frac{m}{\Gamma}\right) .
  \end{align}
With the values used in \Fig{fig:backgroundOmega}, one obtains
$N_\Gamma-N_\uend\simeq 2.8$, which is in good agreement with what can
be observed in this figure. Then, within a few \efolds, radiation
takes over and the radiation-dominated era starts.
In \Fig{fig:backgroundw}, the transparent blue line displays the total
equation-of-state parameter for the background, namely
$w_\mathrm{bg}=(p_\phi+p_\mathrm{f})/(\rho_\phi+\rho_\mathrm{f})$. The
same remarks as in \Fig{fig:backgroundOmega} apply. Initially,
$w_\mathrm{bg}\simeq -1$ and inflation proceeds in the slow-roll
regime, until $w_\mathrm{bg}$ crosses $-1/3$ and inflation
stops. After inflation, $w_\mathrm{bg}$ oscillates, and finally
asymptotes $1/3$ when the transition towards the radiation-dominated
era is completed. In order to factor out the effect of oscillations
and only study their envelope, we also display the averaged value of
$w_\mathrm{bg}$, \ie $\langle w_\mathrm{bg}\rangle$, for two different
time scales of averaging. The orange curve corresponds to
$w_\mathrm{bg}$ convolved with a Gaussian kernel of standard deviation
given by $0.2$ \efold, while the green one follows the same procedure
but with standard deviation $0.1$ \efold. Interestingly, right after
the onset of the oscillatory phase, $\langle w_\mathrm{bg}\rangle$ is
close to zero, which confirms that the background expands on average
as in a matter-dominated era, until the production of radiation
becomes effective.

The behaviour of $\langle w_\mathrm{bg}\rangle$ when radiation is
still subdominant (\ie during inflation and during the first stage of
the oscillating phase) can be described analytically as follows. A
first remark is that \Eq{eq:eomfluidgamma} can be solved exactly,
\begin{align}
\label{eq:rhof:exact}
\rho_{\mathrm{f}}(t)=\frac{\Gamma}{2}\int_{t_\uin}^t  \dot{\phi}^2(\tilde{t})
\left[\frac{a(\tilde{t})}{a(t)}\right]^{3(1+w_{\mathrm{f}})} \dd\tilde{t}\, .
\end{align}
This expression can be cast as a perturbative expansion in
$\Gamma$. At leading order, the integrand should be evaluated with
$\Gamma=0$, \ie using the background dynamics in the absence of the
radiation fluid, which we know how to describe.

During inflation, using the formula given above for the slow-roll
trajectory, one can compute \Eq{eq:rhof:exact} explicitly in terms of
error functions. The resulting expression is not particularly
illuminating so we do not reproduce it here, but we note that if the
initial time is chosen sufficiently far in the past, it converges
to\footnote{This convergence proves that, as mentioned above, after a
  few \efolds~in slow-roll inflation, $\rho_{\mathrm{f}}$ reaches an
  attractor, which implies that our results do not depend on our
  choice of initial time of integration.}
\begin{align}
\label{eq:rhof:inflation:approx}
\rho_{\mathrm{f}} \simeq \frac{m\Mp^2\Gamma}{3}
\sqrt{\frac{\pi}{2\left(1+w_\mathrm{f}\right)}}
\ee^{\frac{3}{2}\left(1+w_\mathrm{f}\right)\left[1-2\left(N-N_\uend\right)\right]}
\mathrm{erfc}\left\lbrace \sqrt{\frac{3}{2}\left(1+w_\mathrm{f}\right)
  \left[1-2\left(N-N_\uend\right)\right]} \right\rbrace\, .
\end{align}
At the end of inflation, $\rho_{\mathrm{f}}$ is therefore of order
$m\Mp^2\Gamma$, hence $\Omega_{\mathrm{f}}$ is of order
$\Gamma/H_\uend$, so radiation can indeed be neglected when the decay
rate is much smaller than the Hubble scale during inflation. For
instance, with the parameter values used in \Fig{fig:backgroundOmega},
\Eq{eq:rhof:inflation:approx} gives $\Omega_\mathrm{f}(t_\uend)\simeq
8.1\times 10^{-4}$ while the numerical integration performed in
\Fig{fig:backgroundOmega} gives $\Omega_\mathrm{f}(t_\uend)\simeq
9.8\times 10^{-4}$, which allows us to check the validity of our
approach (the small difference between these two values is explained
by the fact that the slow-roll approximation breaks down towards the
end of inflation).

During the oscillating phase, in the absence of fluid, as explained
above $\phi(t)\propto \sin(m t) a^{-3/2}$. Plugging this formula into
\Eq{eq:rhof:exact}, and after averaging over the oscillating term, one
obtains
\begin{align} \Omega_{\mathrm{f}} &\simeq
\Omega_{\mathrm{f}}(t_\uend)\ee^{-3
  w_\mathrm{f}\left(N-N_\uend\right)}+ \frac{\Gamma}{12
  H_\uend}\frac{\phi_\uend^2}{\Mp^2} \left\lbrace
\frac{3}{4\left(w_\mathrm{f}-\frac{1}{2}\right)}
\left[\ee^{-\frac{3}{2}\left(N-N_\uend\right)}-\ee^{-3
    w_\mathrm{f}\left(N-N_\uend\right)}\right] \right. \nonumber \\ &
\quad\quad\quad\quad\quad\quad \left.
+\frac{m^2}{3\left(w_\mathrm{f}+\frac{1}{2}\right)H_\uend^2}
\left[\ee^{\frac{3}{2}\left(N-N_\uend\right)}-\ee^{-3
    w_\mathrm{f}\left(N-N_\uend\right)}\right]\right\rbrace\, .
\end{align}
After a few \efolds, if $w_\mathrm{f}>-1/2$, the first term on the
second line is the dominant one, which leads to
\begin{align}
\Omega_{\mathrm{f}} \simeq \frac{1}{18\left(2 w_\mathrm{f}+1\right)}
\frac{\phi_\uend^2 m^2}{\Mp^2 H_\uend^2} \frac{\Gamma}{H}\, .
\end{align}
In a quadratic potential, using the slow-roll formula
$\phi_\uend\simeq \sqrt{2}\Mp$, one has $H_\uend\simeq m/\sqrt{2}$ and
the equation-of-state parameter $w_\mathrm{bg}\simeq w_\mathrm{f}
\Omega_\mathrm{f}$ is given by
\begin{align}
\label{eq:eos_anal}
w_\mathrm{bg} \simeq 
\frac{2w_\mathrm{f}}{9\left(2 w_\mathrm{f}+1\right)}\frac{\Gamma}{H}\, .
\end{align}
Because of the slow-roll violation at the end of inflation, this
formula is expected to provide an accurate description only up to an
overall factor of order one (for instance in $m/H_\uend$), and in
\Fig{fig:backgroundw} one can check that this is indeed the case, see
the inset in particular (the agreement in the case of other fluid
equation-of-state parameters can be checked in \Fig{fig:otherfluids}
below). When $\Gamma$ becomes of order $H$, \ie when $N\sim N_\Gamma$,
the approximation breaks down and \Eq{eq:eos_anal} cannot be trusted
anymore.

\subsection{Perturbations}
\label{subsec:pert}
Having established how the background evolves, we now turn to the
behaviour of the perturbations. Since the equations we started from,
\Eqs{eq:conservationT} and~\eqref{eq:QmuK:to:f}, have a covariant
form, they can be perturbed. As stressed above, this is not the case
of the background equations of motion, \Eqs{eq:eomscalargamma}
and~\eqref{eq:eomfluidgamma}, which explains why these two equations
cannot be used as a starting point, and why it was necessary to
re-derive them from a covariant principle. For more explanations about
this formalism, we refer the interested reader to \Refs{Malik:2004tf,
  Jerome_Lucas_in_prep}. By perturbing \Eq{eq:conservationT}, one
obtains
\begin{align}
\delta\left[\nabla _{\nu}T^{\mu \nu}_{(\alpha)}\right]=\sum_{\beta}
\left[\delta Q^{\mu}_{(\alpha)\rightarrow (\beta)}
-\delta Q^{\mu}_{(\beta)\rightarrow (\alpha)}\right].
\end{align}
In this formula, the perturbed energy transfer $\delta
Q^{\mu}_{(\alpha)\rightarrow (\beta)}$, using Eq.~(\ref{eq:defQ}), can
be written as
\begin{align}
\delta Q^{\mu}_{(\alpha)\rightarrow (\beta)}=\delta Q_{(\alpha)\rightarrow (\beta)}u^{\mu}
+Q_{(\alpha)\rightarrow (\beta)}\delta u^{\mu}
+\delta f^{\mu}_{(\alpha)\rightarrow (\beta)}.
\end{align}
The constraint that the four-vector $f^{\mu}_{(\alpha)\rightarrow
  (\beta)}$ is orthogonal to the Hubble flow must also be satisfied at
the perturbed level, and this leads to $\delta
[f^{\mu}_{(\alpha)\rightarrow (\beta)} u_{\mu}]=0$. As a consequence,
$\delta f^{0}_{(\alpha)\rightarrow (\beta)}=0$ and only $\delta
f_i^{(\alpha)\rightarrow (\beta)} \neq 0$. Since we consider scalar
perturbations, we write $\delta f_i^{(\alpha)\rightarrow
  (\beta)}=\partial_i \delta f_{(\alpha )\rightarrow (\beta)}$ and
work in terms of the function $\delta f_{(\alpha )\rightarrow
  (\beta)}$.

Let us now perturb the gradient of the stress energy tensor for a
scalar field in interaction with a perfect fluid.  At the perturbed
level, the kinetic and potential fictitious fluids associated to
$\phi$ have perturbed energy density and pressure given by
\begin{align}
\delta \rho_{K}^{\rm (gi)}&=\delta p_{K}^{\rm (gi)}
=\frac{\phi'}{a^2}\delta \phi^{\rm (gi)}{}'-
\frac{\phi'{}^2}{a^2}\Phi,
\\
\delta \rho_V^{\rm (gi)}&=-\delta p_V^{\rm (gi)}
=V_{\phi}\delta\phi^{\rm (gi)},
\end{align}
and the perturbed gradient of the stress energy tensor also involves
the velocity potential $v^{\mathrm{(gi)}}_{(\alpha)}$, related to the
spatial component of the perturbed velocity by
$v_{(\alpha),i}^{\mathrm{(gi)}}=\partial_i
v^{\mathrm{(gi)}}_{(\alpha)}$, and the rescaled velocity
$\varsigma_{(\alpha)}^{\mathrm{(gi)}}$ defined by
$\varsigma_{(\alpha)}^{\mathrm{(gi)}}\equiv
[\rho_{(\alpha)}+p_{(\alpha)}]v_{(\alpha)}^{\rm (gi)}$, \bea
v_{K}^{\rm (gi)}=-\frac{\delta \phi^{\rm (gi)}}{\phi'}, \quad
\varsigma_{K}^{\rm (gi)}=-\frac{\phi'}{a^2}\delta \phi^{\rm (gi)}.
\eea Notice that we do not need to specify $v_{V}^{\rm (gi)}$ since it
does not appear in the equations. In these expressions, as already
mentioned, the superscript ``$(\mathrm{gi})$'' means that the
corresponding quantity is gauge-invariant and coincides with its value
in the longitudinal gauge.  At the perturbed level, the
energy-momentum transfer coefficients between the kinetic and
potential fluids are given by
\begin{align}
a\delta Q_{K\rightarrow V}&=-V_{\phi}\delta \phi^{\rm (gi)}{}'
+V_{\phi}\phi'\Phi-V_{\phi\phi}\phi'
\delta\phi^{\rm (gi)}, \quad
a\delta Q_{V\rightarrow K}=0, \\
\delta f_{K\rightarrow V}&=
\delta f_{V\rightarrow K}=0.
\end{align}
As will be shown below, these formulas are indeed needed to recover
the standard equation of motion for the scalar field fluctuation (\ie
the equation of motion in absence of coupling with a fluid, for which
a Lagrangian formulation of the theory exists and the equation of
motion is well prescribed).  Regarding the interaction between the
kinetic and potential fluids on one hand, and the perfect fluid on the
other hand, we have from perturbing \Eq{eq:QmuK:to:f}
\begin{align}
\label{eq:deltaQ:K:f}
\delta Q_{K\rightarrow \mathrm{f}}& =-\Gamma\delta \rho_{K}^{\mathrm{(gi)}}, \quad
\delta Q_{\mathrm{f}\rightarrow K}=
\delta Q_{V\rightarrow \mathrm{f}} =\delta Q_{\mathrm{f}\rightarrow V}=0,
\end{align}
and
\begin{align}
\label{eq:delta:f:K:f}
  \delta f_{K\rightarrow \mathrm{f}}&=a\Gamma 
  \left[v_\mathrm{tot}^{\mathrm{(gi)}}
  -v_K^{\mathrm{(gi)}}\right]\rho_K, \quad \delta f_{\mathrm{f}\rightarrow K}= 
  \delta f_{V\rightarrow \mathrm{f}}=\delta f_{\mathrm{f}\rightarrow V}=0,
\end{align}
where the total velocity $v^{\rm (gi)}_\mathrm{tot}$ is defined by the
following expression
\begin{align}
v^{\rm (gi)}_\mathrm{tot}=\frac{1}{\rho+p}\sum _{\alpha}
\left[\rho_{(\alpha)}+p_{(\alpha)}\right]v_{(\alpha)}^{\rm (gi)},
\end{align}
with $\rho$ and $p$ the total energy density and pressure.

\begin{figure}[t]
\begin{center}
  \includegraphics[width=1.0\textwidth]{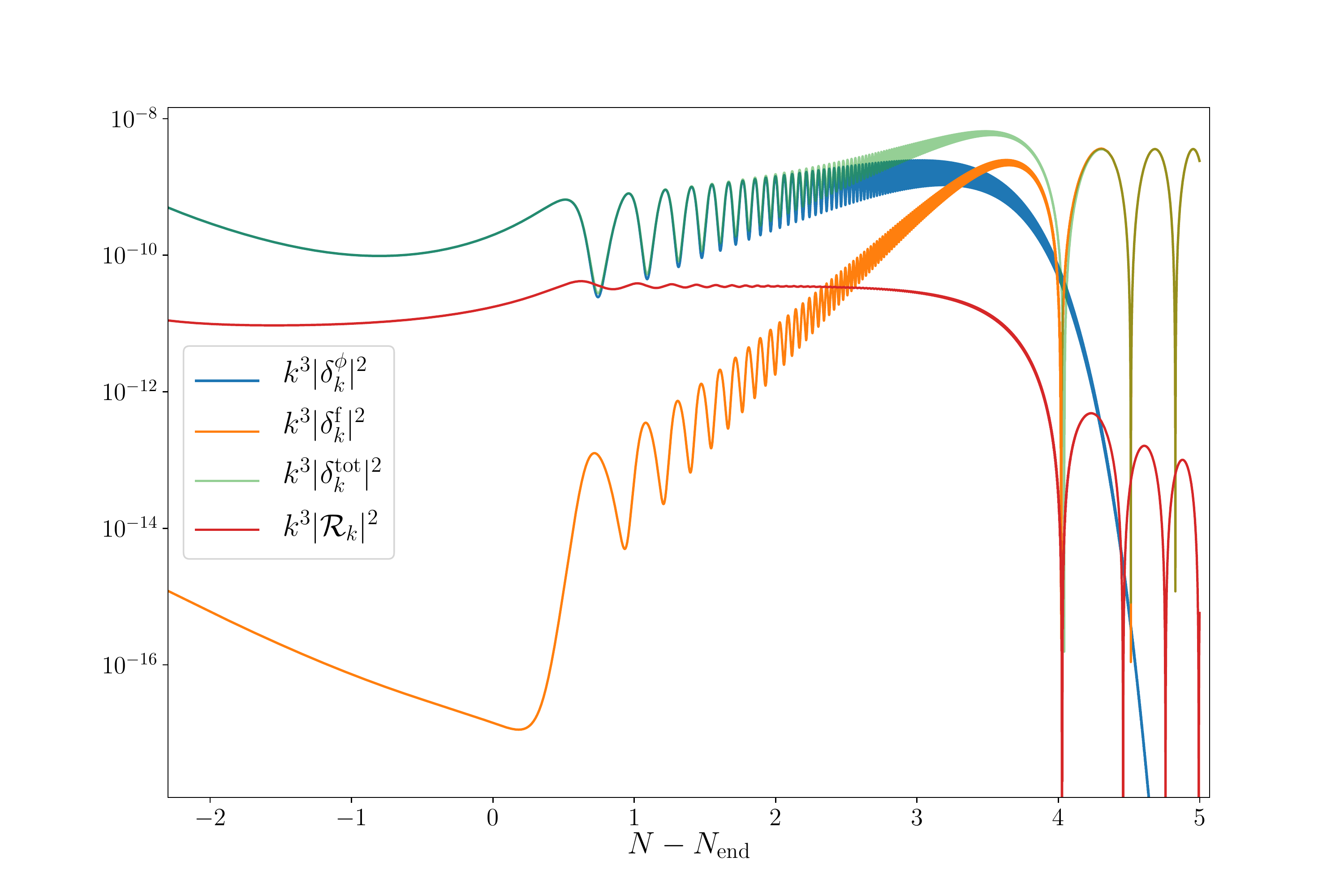}
\caption{Evolution of the modulus of the gauge invariant perturbative
  quantities $\delta_k^\phi$ (inflaton density contrast, blue line),
  $\delta_k^\mathrm{f}$ (radiation density contrast, orange line),
  $\delta_k^\mathrm{tot}$ (total, \ie scalar field plus radiation,
  density contrast, green line) and $\mathcal{R}_k$ (comoving
  curvature perturbation, red line) as a function of the number of
  \efolds, in the same setup as the one displayed in the previous
  figures. Soon after perturbative reheating becomes effective (which,
  according to the discussion around \Fig{fig:backgroundOmega}, occurs
  when $N_\Gamma-N_\uend\simeq 2.8$), the scalar field density
  contrast decreases, the curvature perturbation stops being constant
  and decreases as well, hence the total density contrast stops
  increasing, which signals the end of the instability.}
\label{fig:perturbations}
\end{center}
\end{figure}

Endowed with these definitions and assumptions one can then derive the
perturbed equations of motion. For the scalar field, one obtains the
perturbed Klein-Gordon equation
\begin{align}
\delta \phi^{\mathrm{(gi)}}{}''+2{\cal H}\delta \phi^{\mathrm{(gi)}}{}'
  +\frac{a\Gamma}{2}\delta \phi^{\mathrm{(gi)}}{}'
  -\nabla^2\delta \phi^{\mathrm{(gi)}}{}+a^2 V_{\phi \phi}
  \delta \phi^{\mathrm{(gi)}}=4\phi'\Phi'-2a^2V_\phi\Phi
  -\frac{a\Gamma}{2}\phi'\Phi.
\end{align}
For the perfect fluid, one has two equations, namely the time and space
components of the conservation equation, yielding an equation for the
perturbed energy density and the perturbed velocity respectively, which read
\begin{align}
  & \delta \rho^{\mathrm{(gi)}}{}'+3{\cal H}(1+w_\mathrm{f})\delta \rho^{\mathrm{(gi)}}
  -3(1+w_\mathrm{f})\rho \Phi'
  +(1+w_\mathrm{f})\rho \nabla^2 v^{\mathrm{(gi)}}
\nonumber \\ &
  -\frac{\Gamma}{a}\left[\phi'\delta \phi^{\mathrm{(gi)}}{}'
  -\frac12\phi'^2\Phi\right]=0, \\
  &\varsigma^{\mathrm{(gi)}}{}'+4{\cal H}\varsigma^{\mathrm{(gi)}}+\rho(1+w_\mathrm{f})\Phi
  +w_\mathrm{f}\delta \rho^{\mathrm{(gi)}}{}+\frac{\Gamma}{2a}
  \phi'\delta \phi^{\mathrm{(gi)}}=0.
\end{align}
One also needs an equation to track the evolution of the
Bardeen potential and this is provided by the perturbed Einstein
equations,
\begin{align}
  \Phi'=-{\cal H}\Phi-\frac{a^2}{2\Mp^2}\left[-\frac{1}{a^2}
  \phi'\delta \phi^{\mathrm{(gi)}}+\varsigma^{\mathrm{(gi)}}{}\right].
  \end{align}

In \Fig{fig:perturbations}, we have numerically integrated the above
equations using the same parameters as in \Figs{fig:backgroundOmega}
and~\ref{fig:backgroundw} and for the mode $k/a_\uini=0.002 \Mp$, the
physical wavelength of which is displayed in \Fig{fig:scale}. The
solid blue line in \Fig{fig:perturbations} represents the scalar field
density contrast $k^3\vert \delta_{\bm k}^{\phi}\vert^2=k^3\vert
\delta \rho_{\phi,\bm k}^{\mathrm{(gi)}}/\rho_\phi\vert^2$, the solid
orange line corresponds to the radiation fluid density contrast
$k^3\vert \delta_{\bm k}^{\mathrm{f}}\vert^2=k^3\vert \delta
\rho_{\mathrm{f},\bm k}^{\mathrm{(gi)}}/\rho_\mathrm{f}\vert^2$, while
the green line is the total density contrast $k^3\vert \delta
^\mathrm{tot}_{\bm k}\vert^2=k^3\vert [\delta \rho_{\phi,\bm
    k}^{\mathrm{(gi)}}+\delta \rho_{\mathrm{f},\bm
    k}^{\mathrm{(gi)}}{}]/(\rho_\phi+\rho_\mathrm{f})\vert^2$. When
the mode enters the instability band around $N-N_\mathrm{end}\simeq
0.5$ \efold, we see that the scalar field density contrast grows and
one can check that this growth is proportional to the scale factor
$a(t)$. This is a first consistency check. Originally, this growth was
derived from an analysis based on the Mathieu-like equation for the
Mukhanov-Sasaki variable, see \Ref{Jedamzik:2010dq}. Here, we recover
it using the conservation equations. We also notice that, initially,
the total density contrast is equal to the scalar field density
contrast which is of course expected since the production of radiation
has not yet started in a sizeable way. When the amount of radiation
starts being substantial, the two density contrasts become different
as revealed by the fact that the green and blue curves separate. Then,
the scalar field density contrast strongly decreases and becomes
quickly negligible. This means that the total density contrast is
given by the radiation density contrast and we see that, when the
transition is completed, it stays constant. In
\Fig{fig:perturbations}, we have also represented the comoving
curvature perturbation ${\cal R}_{\bm k}=\Psi_{\bm k}-aH v^{\rm
  (gi)}_{\mathrm{tot},\bm k}$ with the red line.  At the onset of the
instability phase, it is, as expected from the above analysis,
constant, and then it decreases as expected for sub-sonic
perturbations in a radiation-dominated universe.

The main conclusion of this analysis is a confirmation that
perturbative reheating effects do not destroy the metric preheating
instability, since the instability stops only when, at the background
level, the radiation fluid dominates the energy budget of the
universe.  The tiny amount of radiation that is initially present is
not sufficient to blur the narrow-resonance regime and to remove the
system from the first, and very thin, instability band of the Mathieu
equation chart. Notice that this supports the treatment of
\Ref{Martin:2019nuw} where the instability was simply stopped at the
time when the universe becomes radiation dominated. This also
demonstrates the robustness of the results obtained in
\Ref{Jedamzik:2010dq} and the generic, unavoidable presence of an
instability in single-field models of inflation at small scales.

\begin{figure}[t]
\begin{center}
  \includegraphics[width=0.49\textwidth]{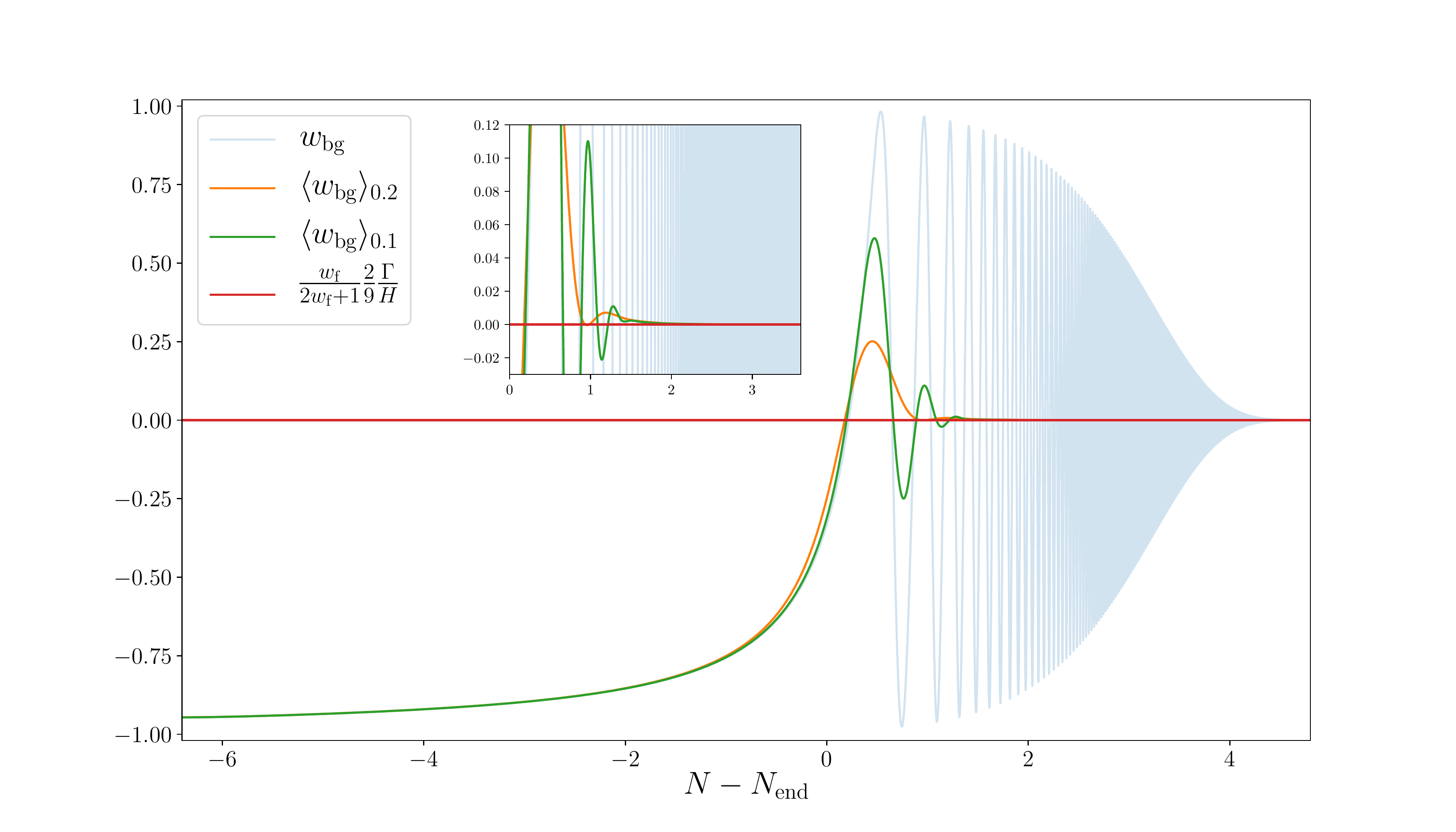}
  \includegraphics[width=0.49\textwidth]{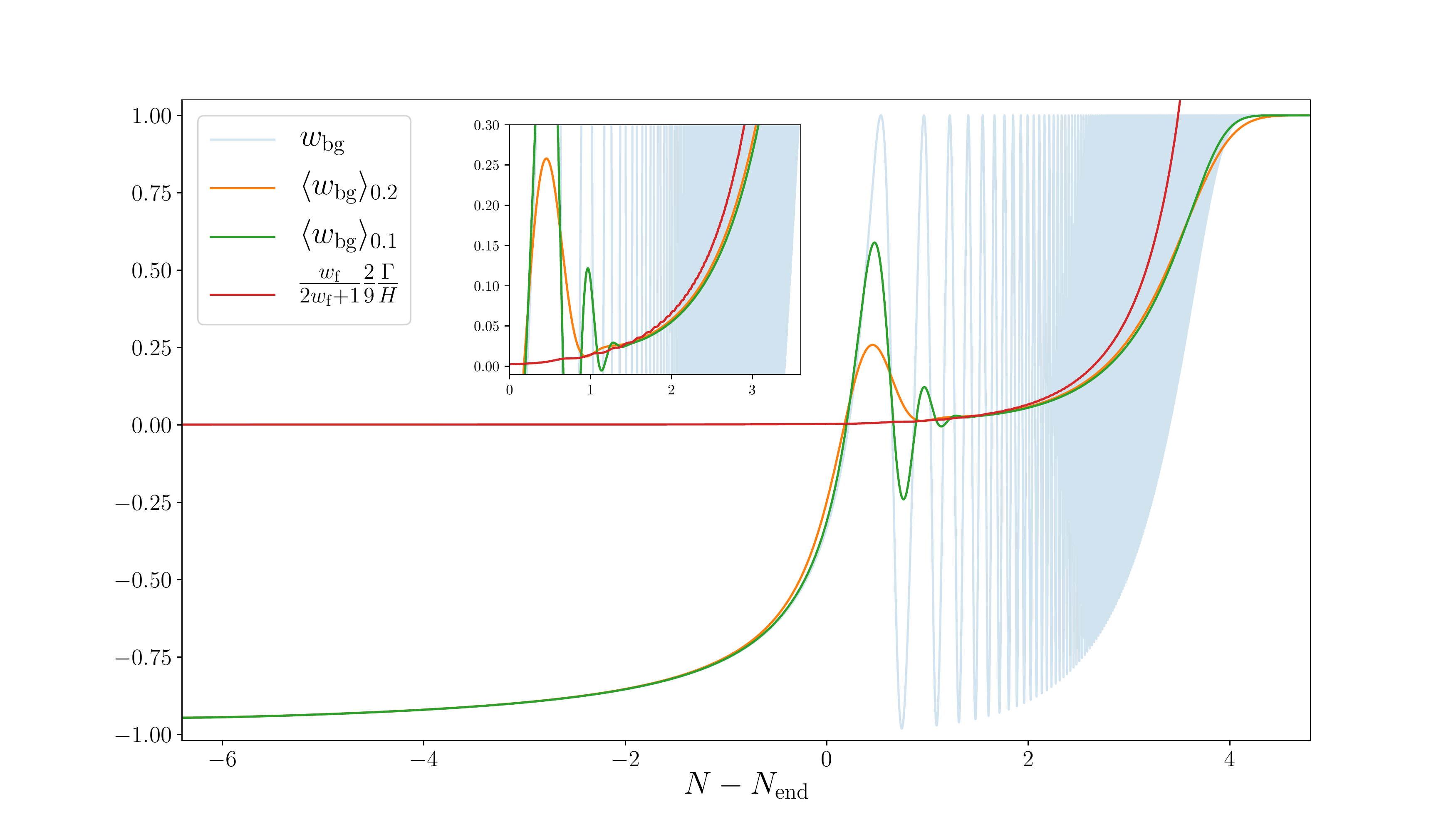}
    \includegraphics[width=0.49\textwidth]{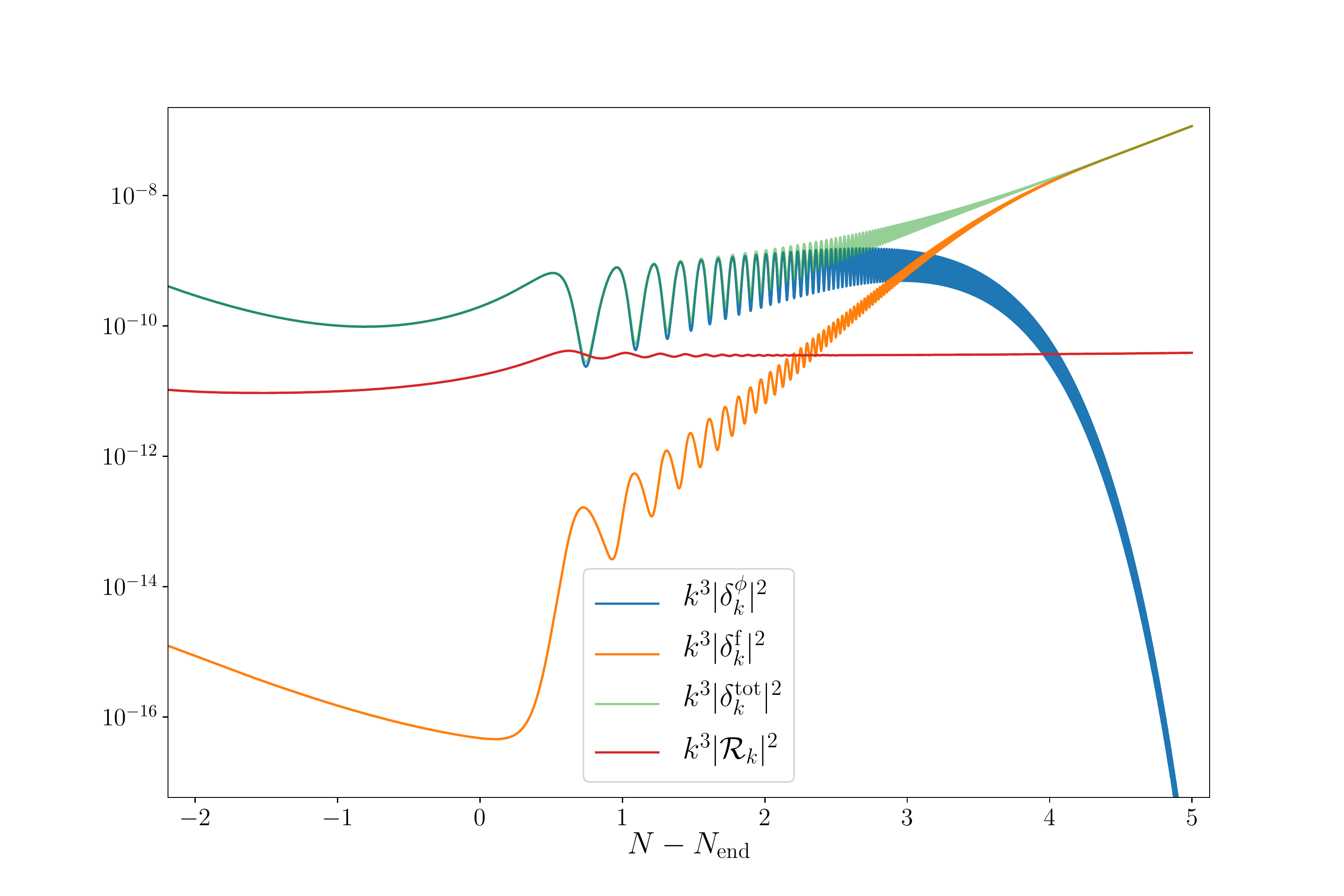}
  \includegraphics[width=0.49\textwidth]{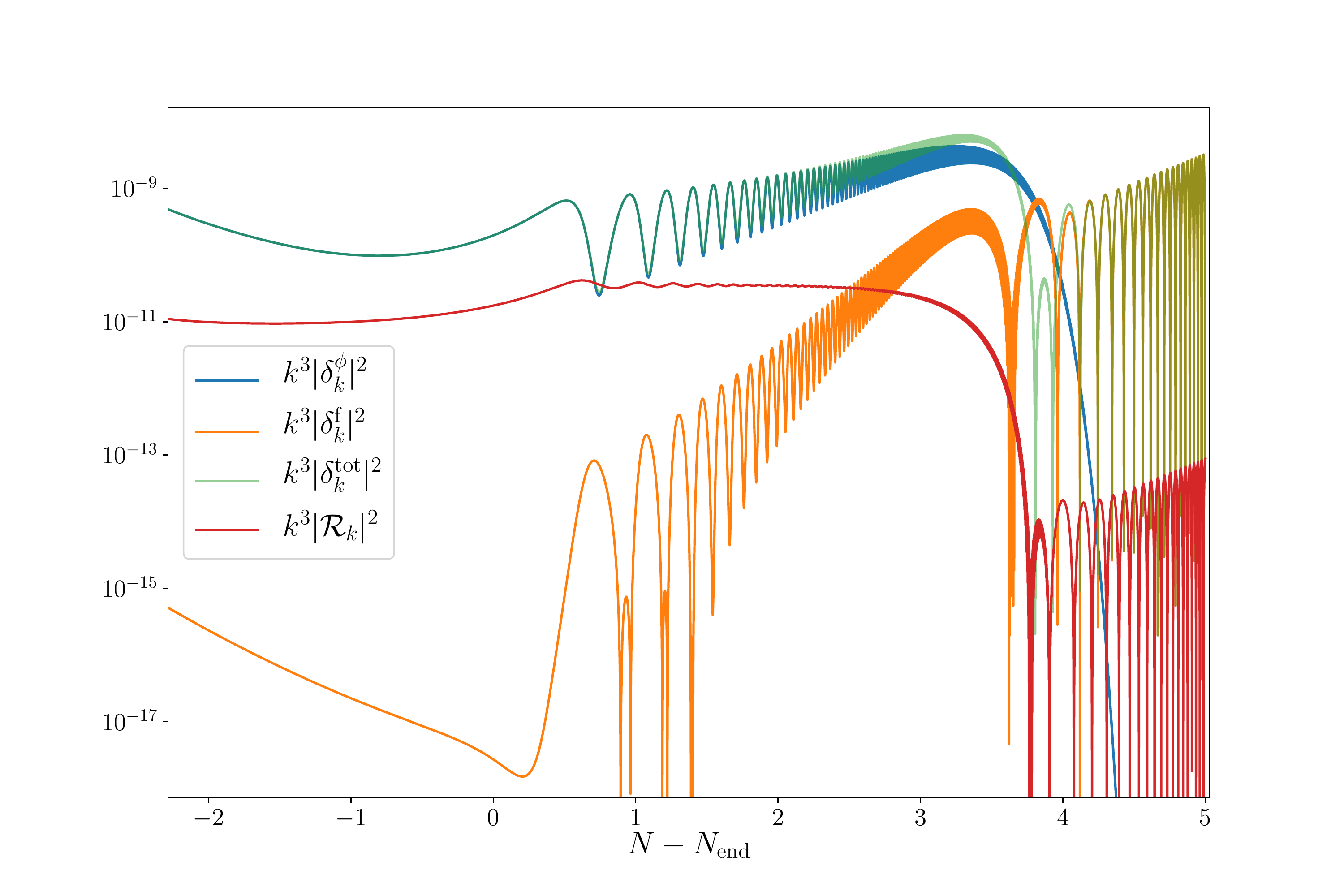}
\caption{Time evolution of the background equations-of-state
  parameters (upper panels), with the insets zooming in the regime of
  validity of the analytical approximation~\eqref{eq:eos_anal}, as
  well as scalar perturbations (lower panels), as a function of the
  number of \efolds, in the cases of decay into pressureless matter
  $w_\mathrm{f}=0$ (left panels) and into a stiff fluid
  $w_\mathrm{f}=1$ (right panels). Apart from the value of
  $w_\mathrm{f}$, the setup and parameter values are the same as in
  all previous figures.}
\label{fig:otherfluids}
\end{center}
\end{figure}

Another way to test this robustness is to study whether the above
conclusion is still valid when the inflaton decays into a fluid with
an equation of state that differs from the one of radiation. We have
therefore considered two additional cases corresponding to a decay
into a fluid with $w_\mathrm{f}=0$ (pressureless matter) and a decay
into a fluid with $w_\mathrm{f}=1$ (stiff matter). The results are
displayed in \Fig{fig:otherfluids} and confirm that our description of
the instability generalises to arbitrary equation-of-state parameters
$w_\mathrm{f}$. On the upper panels, we show the total equations of
state $w_{\mathrm{bg}}$ and their averaged values, as well as the
analytical approximation \Eq{eq:eos_anal}, as a function of the number
of \efolds. One verifies that the equation-of-state parameter indeed
asymptotes $w_{\mathrm{bg}}=0$ (left panel) and $w_{\mathrm{bg}}=1$
(right panel) at late time. On the lower panels, we have displayed the
time evolution of the density contrasts and of the curvature
perturbation. The growth $\delta_k\propto a$ is still observed until
the universe is dominated by the fluid,\footnote{In the case where the
  decay product is a pressureless fluid, the growth $\delta_k\propto
  a$ still continues afterwards for all scales. In the case
  where $w_\mathrm{f}=1$, stiff fluid density fluctuations also grow
  like $\delta_k\propto a$ on sub-Hubble scales, see the relation
  above \Eq{eq:w:eff:sub_sonic}.} regardless of its equation of state.

\subsection{Radiative decay and PBH formation from metric preheating}
\label{subsec:pbh}
In the covariant description developed in \Sec{sec:radiativedecay},
two fluids were necessary to fully describe the scalar field
fluctuations. This shows that cosmological inhomogeneities of a scalar
field and of a perfect fluid are a priori two very different physical
systems, featuring different properties. It is therefore rather
intriguing that, during the oscillatory phase, the averaged equation
of state is the one of pressureless matter, and that inside the
instability band, the density contrast behaves as the one of
pressureless matter too, since it grows linearly with the scale
factor.

The formation of primordial black holes has mostly been studied in the
context of perfect fluids, so if this correspondence between an
oscillating scalar field and a pressureless perfect fluid does hold
(and even in the presence of additional radiation), it would have
important practical consequences~\cite{Carr:2018nkm} for studying the
production of PBHs from the metric preheating instability. This is
why, in this section, we compare more carefully the behaviour of the
cosmological perturbations of the system at hand with those of a
single perfect fluid sharing the same equation-of-state parameter.

A key concept in this comparison is the one of the equation of state
``felt'' by the perturbations, if they are interpreted as
perturbations of a single perfect fluid. We start by recalling the
behaviour of the density contrast for a perfect fluid with a given
equation-of-state parameter $w$. This will allow us to extract the
equation-of-state parameter from the time dependence of the density
contrast, and to apply this formula to the system studied in
\Sec{sec:radiativedecay} in order to derive the effective ``equation
of state'' felt by the density perturbations. We will then compare it
with the equation of state of the background.

In order to implement this program, a remark is in order regarding the
definition of the density contrast. So far, we have worked in terms of
the density contrast $\delta^{\mathrm{(gi)}}$ (noted $\delta
_{\mathrm{g}}$ in \Ref{Bardeen:1980kt}),
which consists in measuring the
energy density relative to the hypersurface which is as close as
possible to a ``Newtonian'' time slicing. However, for a single
perfect fluid, this density contrast usually stays constant at large
scales and, as a consequence, cannot be used as a tracer of the
equation-of-state parameter. Fortunately, as is well-known, there are
other possible definitions, in particular $\delta_\mathrm{com}$ (noted
$\delta_{\mathrm{m}}$ in \Ref{Bardeen:1980kt}), which measures the
amplitude of energy density from the point of view of matter, and
corresponds to the density contrast in the comoving-orthogonal
gauge. The behaviour of $\delta_\mathrm{com}$ does depend on $w$ on
large scales and, therefore, it is a useful quantity for our
purpose. The relationship between $\delta ^\mathrm{(gi)}$ and
$\delta_\mathrm{com}$ is given by 
\begin{align}
\delta^\mathrm{(gi)}=\delta
_\mathrm{com}-\frac{\rho'}{\rho}v^{(\mathrm{gi})}=\delta
_\mathrm{com} \left[1+ 3 \frac{a^2 H^2 }{k^2}
  \left(1+ \frac{\Phi^\prime}{ a H \Phi} \right) \right],
\end{align}
which shows that, although they behave differently on super-Hubble scales, their
evolution is identical on small scales. To prove this relation, we have
used that the density contrast $\delta_\mathrm{com}$ is related to
the Bardeen potential through the Poisson
equation~\cite{Bardeen:1980kt}
\begin{align}
  \delta _\mathrm{com}=-\frac{2k^2\Mp^2}{a^2\rho}\Phi\, .
  \end{align}
If the space-time expansion is driven by a perfect fluid with constant
equation-of-state parameter $w$, the energy density scales as
$\rho=\rho_\uend(a_\uend/a)^{3(1+w)}$, which leads to
\begin{align}
\label{eq:deltam:Phi:a}
  \delta_\mathrm{com}=\delta_\mathrm{com}^\mathrm{end}
  \left(\frac{a}{a_\uend}\right)^{1+3w}\frac{\Phi}{\Phi_\uend}, 
  \end{align}
where the Bardeen potential follows the equation of
motion~\cite{Mukhanov:1990me}
\begin{align}
  \frac{\dd^2 }{\dd (k\eta)^2}\left[(k\eta)^\nu\Phi\right]
  +\frac{2}{k\eta}\frac{\dd }{\dd k\eta}\left[(k\eta)^\nu\Phi\right]
  +\left[w- \frac{\nu(\nu+1)}{\left(k\eta\right)^2}\right](k\eta)^\nu\Phi=0\, ,
\end{align}
with $\nu=2/(1+3w)$. The solution to this equation is given by  
\begin{align}
\label{eq:Phi:perfect:fluid}
\Phi_{\bm k}=(wk\eta)^\alpha\left[A_{\bm
    k}J_{\alpha}(wk\eta)+B_{\bm k}J_{-\alpha}(wk\eta)\right],
\end{align}
with $\alpha=-(5+3w)/[2(1+3w)]$, $J_{\alpha}$ being a Bessel function
and $A_{\bm k}$, $B_{\bm k}$ two integration constants fixed by the
initial conditions. The behaviour of this solution depends on whether
$\vert wk\eta\vert \ll 1$ or $\vert wk\eta\vert \gg 1$, \ie on whether
the mode wavelength is larger or smaller than the sound horizon $w/H$.

On super-sonic scales, $\vert wk\eta\vert \ll 1$, the Bessel functions
can be expanded according to $J_\alpha(z)\propto z^\alpha$. Since
$\alpha<0$ for $w>-1/3$, \Eq{eq:Phi:perfect:fluid} features a constant
mode and a decaying mode. The Bardeen potential thus asymptotes to a
constant, and $\delta_\mathrm{com} \propto a^{1+3w}$, see
\Eq{eq:deltam:Phi:a}. If $w=0$, then $\delta_\mathrm{com}\propto a$,
which is a well-known result.

On sub-sonic scales, $\vert wk\eta\vert \gg 1$, the Bessel functions
can be expanded according to $J_\alpha(z) \simeq \sqrt{2/(\pi
  z)}\cos[z-\pi(1+2\alpha)/4]$. This leads to $\delta_\mathrm{com}\simeq
a^{-1/2+3w/2}\cos[wk\eta-\pi(1+2\alpha)/4]$. The density contrast thus
oscillates as a result of the competition between gravity and
pressure, and compared to the super-sonic case, the overall amplitude
also scales differently with the scale factor.  One also notices that
this formula cannot be applied if $w=0$. Indeed, in that case, the
argument of the Bessel functions vanishes. Physically, if $w=0$, there
is no sound horizon anymore (since the pressure vanishes), and all
scales are ``super-sonic'' by definition.
\begin{figure}[t]
\begin{center}
  \includegraphics[width=0.8\textwidth]{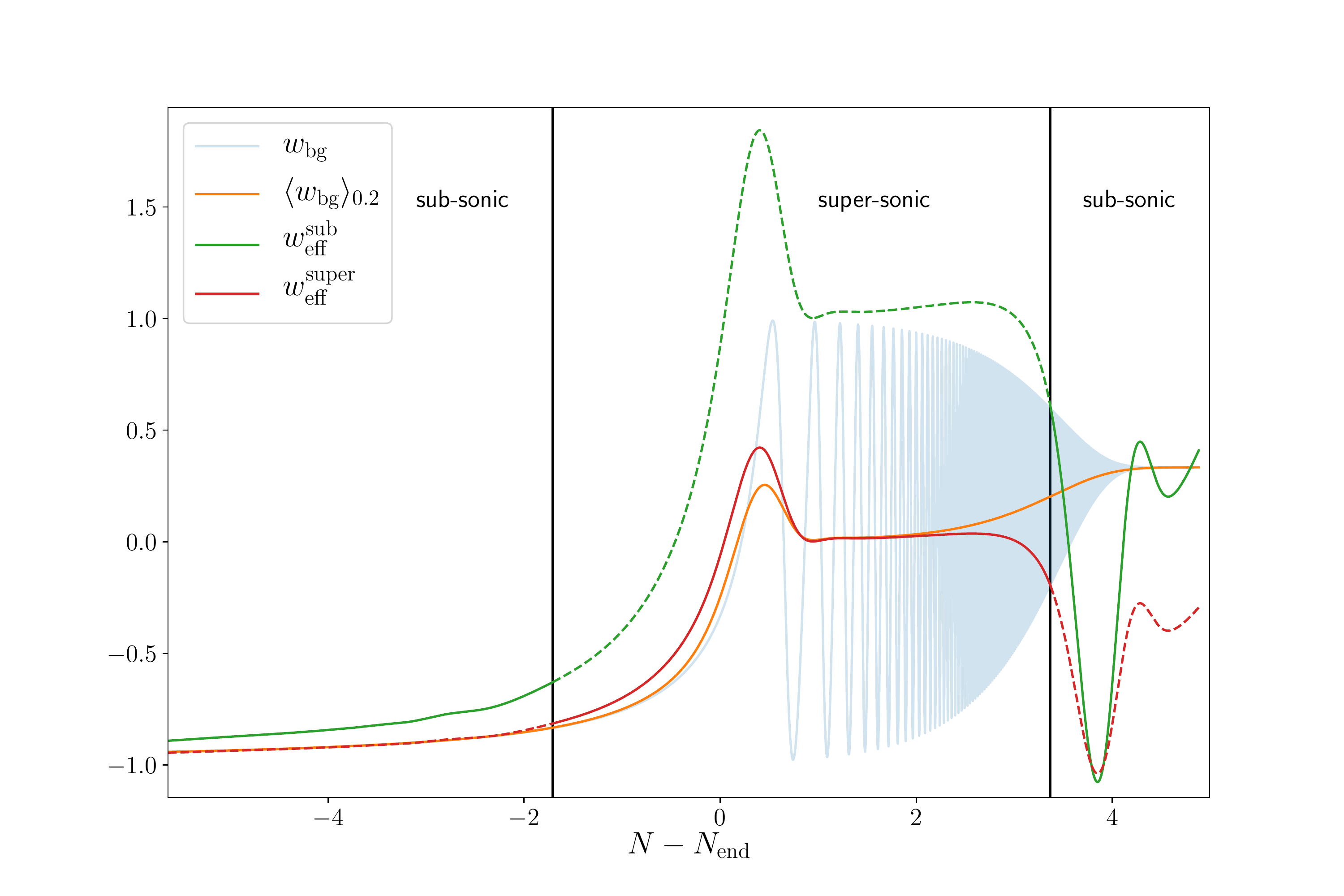}
  \caption{Effective equation-of-state parameters for the
    perturbations as a function of the number of
    \efolds. 
    When the mode $k$ is super-sonic, its effective equation-of-state parameter is given by $w_\mathrm{eff}^\mathrm{super}$ (red line), and respectively by $w_\mathrm{eff}^\mathrm{sub}$ (green line) when it is sub-sonic. In order to facilitate the reading of the
      figure, the effective equation-of-state parameters are displayed with dashed lines in the regimes where they are not relevant.
 The instantaneous background equation of state (transparent blue line)
    and its averaged value (orange line) are also represented for
    comparison. When $\left\langle w_\mathrm{bg} \right\rangle\approx
    0$, the sound horizon is very small and the super-sonic effective
    equation of state (red line) is the relevant one. As expected, it
    is close to $0$. However when $\left\langle w_\mathrm{bg}
    \right\rangle$ starts to depart from zero, since the physical mode
    $k/a$ is within the Hubble radius (see \Fig{fig:scale}), the
    sub-sonic equation of state (green line) becomes the relevant one
    and, as expected, it quickly converges to $1/3$. Note however that
    between these two asymptotic regimes, the effective equation of
    state for the perturbation does not match the one of the
    background.  }
\label{fig:effw}
\end{center}
\end{figure}

These two limiting expressions of the density contrast can be used to
define an effective equation-of-state parameter ``felt'' by the
perturbations. Since $\delta_\mathrm{com} \propto a^{1+3w}$ on
super-sonic scales, we define
\begin{align}
  w_\mathrm{eff}^\mathrm{super}\equiv \frac{1}{6}\frac{\mathrm{d}
    \ln \left(k^3\left\langle \delta_\mathrm{com}^2 \right\rangle\right)}{\mathrm{d}\ln a}
  -\frac{1}{3}\, ,
  \end{align}
where $\langle \cdot \rangle$ stands for time averaging over possible
background oscillations. On sub-sonic scales, $\delta_\mathrm{com}\simeq
a^{-1/2+3w/2}\cos[wk\eta-\pi(1+2\alpha)/4]$, so we introduce
\begin{align}
\label{eq:w:eff:sub_sonic}
w_\mathrm{eff}^\mathrm{sub}\equiv\frac{1}{3} \frac{\dd \ln \left(k^3
  \left\langle \delta_\mathrm{com}^2 \right\rangle\right)}{\dd \ln a }+\frac{1}{3}\, .
  \end{align}
Which of these two effective equations of state is relevant depends on
whether the mode $k$ is sub-sonic and super-sonic. In \Fig{fig:effw},
we display these two quantities, $w_\mathrm{eff}^\mathrm{super}$ and
$w_\mathrm{eff}^\mathrm{sub}$, from the value of $\delta_\mathrm{com}$
numerically obtained as in the previous figures, and compare them with
the (averaged) equation-of-state parameter of the background. In all
cases, the time averaging is performed with a Gaussian kernel of
constant standard deviation given by $0.2$ \efolds. Let us also stress
again that, on sub-Hubble scales, the density contrast in the
comoving-orthogonal gauge, $\delta_\mathrm{com}$, coincides with the
one in the longitudinal gauge displayed in \Figs{fig:perturbations}
and~\ref{fig:otherfluids}.

During the first oscillations, the equation-of-state parameter
vanishes (on average) in the background, and recalling that all modes
are super-sonic for a vanishing equation-of-state parameter, one can
check that the relevant equation of state,
$w_\mathrm{eff}^\mathrm{super}$, indeed vanishes, and that the red and
orange curves in \Fig{fig:effw} are indeed close. This however lasts
for a few \efolds~only, after which neither of the effective equations
of state correctly reproduces the behaviour of the (averaged) equation
of state of the background. In addition, for the sub-sonic scales that
lie inside the instability band, $\langle w_\mathrm{bg}\rangle$ does
not coincide at all with $w_\mathrm{eff}^\mathrm{sub}$ during the
oscillating phase until radiation strongly dominates the universe
content and both converge to $1/3$. 
Therefore, despite the fact that
the inflaton background effectively behaves as pressureless matter on
average, and that its decay product is a perfect fluid, the
perturbations of the system are not those of perfect fluids. This
confirms that the system made of a decaying, oscillating scalar field
has different behaviour from a pure perfect fluid, and cannot be
simply modelled as such.

Let us note that this fundamental difference is even more striking in
the case where the inflaton potential is quartic close to its minimum,
since in that case the correspondence between the inflaton
perturbations and those of a perfect fluid with the same background
equation of state breaks down even in the absence of inflaton
radiative decay. As shown in \Ref{Jedamzik:2010dq} indeed, while
$\langle w_\mathrm{bg} \rangle =1/3$ in such a case, the instability
of metric preheating is still present, and the density contrast grows
even faster than that of pressureless matter (namely, exponentially
with the scale factor) in the instability band, while the density
contrast for a perfect fluid having $w=1/3$ is constant on sub-sonic
scales.
  
As mentioned above, this implies that, in order to study the
production of PBHs that arises from the increase of the density
contrast in the instability band, one cannot rely on techniques
developed for perfect fluids. In \Ref{Carr:2018nkm} for instance, it
was used that an overdensity of a perfect fluid with constant
equation-of-state parameter $w$ collapses into a black hole if it
exceeds the critical density contrast\footnote{The criterion for
    PBH formation is expressed in terms of the density contrast rather
    than curvature perturbation, the latter being affected
    by environmental effects~\cite{Yoo:2018kvb}, see also
    \Ref{Young:2014ana}.}~\cite{Harada:2013epa, Harada:2017fjm}
\begin{align}
  \label{eq:deltac:perfect:fluid}
  \delta _\mathrm{c}=\frac{3(1+w)}{5+3w}
  \sin^2\left(\frac{\pi \sqrt{w}}{1+3w}\right)\, ,
\end{align}
in which $w$ was replaced with $w \sim \Gamma/H$ [see
  \Eq{eq:eos_anal}]. If $w=0$, \Eq{eq:deltac:perfect:fluid} indicates
that any local overdensity ends up forming a black hole, which is
indeed the case in the absence of any pressure force. The analysis of
\Ref{Carr:2018nkm} thus suggests that what limits the formation of
PBHs from the instability of metric preheating is the presence of
(even small amounts of) radiation, which provide a non-vanishing value
to the equation-of-state parameter, and hence to
$\delta_\uc$. However, the results of the present work cast some doubt
on such a treatment since we showed that an oscillating scalar field
decaying into a radiation fluid cannot be treated as a collection of
perfect fluids at the perturbed level [furthermore, the background
  equation of state for such a system is strongly time dependent, see
  \Eq{eq:eos_anal}, while \Eq{eq:deltac:perfect:fluid} only applies to
  constant equation-of-state parameters].

In \Ref{Martin:2019nuw}, the formation of PBHs from the overdensities
of an oscillating scalar field was studied in the context of metric
preheating, and it was found that what limits the formation of PBHs is
rather the fact that the instability does not last for ever, since it
stops when radiation takes over. Indeed, although it is true that any
overdensity inside the instability band develops towards forming a
black hole, the amount of time needed for a black hole to form depends
on (and decreases with) the initial value of the density contrast. By
requiring that it takes less time than what is available before the
complete inflaton decay (which, as we have established in
\Sec{subsec:pert}, signals the end of the instability phase that is
otherwise not affected by the presence of radiation being produced),
one obtains a lower bound on the density contrast, which however has
nothing to do with \Eq{eq:deltac:perfect:fluid}.

\section{Conclusions}
\label{sec:conclusions}

Preheating effects are often believed to be observationally irrelevant
in single-field models of inflation. Although this is true at large
scales, where the curvature perturbation is merely conserved, the
situation is different at small scales, namely those leaving the
Hubble radius a few \efolds~before the end of inflation. Such scales
are subject to a persistent instability proceeding in the
narrow-resonance regime~\cite{Jedamzik:2010dq}, which causes the
density contrast to grow, leading to various possible effects such as
early structure formation or even PBHs
formation~\cite{Martin:2019nuw}.

In contrast to the case of background preheating, where the
narrow-resonance regime is irrelevant since, in a time-dependent
background, the system spends very little time in the thin instability
band and the resonance effects are wiped out, in the metric preheating
case, the presence of the instability is actually caused by cosmic
expansion itself (see \Fig{fig:scale}). This is the reason why this mechanism
is both atypical and very efficient.

This fact was known to be true~\cite{Jedamzik:2010dq} only if the
inflaton is uncoupled to other degrees of freedom. However, in order
for reheating to proceed, the inflaton field must decay into
radiation, and the goal of this paper was to determine whether this
decay could spoil the instability. Using the formalism of cosmological
perturbations in the presence of interactions between fluids, we have
shown that it is not the case, and that the growth of the density
contrast inside the instability band remains unaffected until the
radiation fluid dominates the universe content.

We have also stressed that there is a fundamental difference between
the cosmological perturbations of an oscillating scalar field and
those of a perfect fluid, and that techniques developed to study the
formation of PBHs from perfect fluid overdensities cannot be applied
to the present context. Instead, a dedicated analysis such as the one
of \Ref{Martin:2019nuw} must be performed. Our results have confirmed
that the presence of radiation can simply be ignored until it comes to
dominate the energy budget, thus stopping the instability.

The results of this work therefore confirm that the instability of
metric preheating is unavoidable in single-field models of inflation,
since it only requires an oscillating scalar field in a cosmological
background, which is the state of the universe at the end of most
inflationary models, and given that it is robust against perturbative decay of this field.

\begin{acknowledgments}
T.~P. acknowledges support from a grant from the Fondation CFM pour la
Recherche in France as well as funding from the Onassis Foundation -
Scholarship ID: FZO 059-1/2018-2019, from the Foundation for Education
and European Culture in Greece and the A.G. Leventis Foundation.
\end{acknowledgments}
\bibliographystyle{JHEP}
\bibliography{instab}
\end{document}